\documentclass[sigconf,balance=false]{acmart}

\usepackage{popets}

\usepackage[normalem]{ulem}
\usepackage{hyperref}
\usepackage{makecell}
\usepackage{pifont}
\usepackage{rotating}
\usepackage{tikz}
\useunder{\uline}{\ul}{}
\usepackage{subcaption}
\usepackage{xspace}
\usepackage{enumitem}
\usepackage{listings}
\usepackage{pifont}
\usepackage{float}
\usepackage{colortbl}
\usepackage{xcolor}
\usepackage{booktabs}
\usepackage{geometry}
\usepackage{multirow}
\usepackage{hyperxmp}
\usepackage{graphicx}
\usepackage{longtable}
\usepackage{array}

\newif\ifreview

\newcommand{\rotrot}[1]{\rotatebox{90}{#1}}

\usepackage{fp}
\newcommand{\maxnum}{100.00}
\newlength{\maxlen}

\newcommand{\databarfull}[2][red!30]{%
  \settowidth{\maxlen}{\maxnum}%
  \addtolength{\maxlen}{\tabcolsep}%
  \FPeval\result{round(#2/\maxnum:4)}%
  \FPeval\toprint{#2}%
  \rlap{\color{#1}\hspace*{-.5\tabcolsep}\rule[-.05\ht\strutbox]{\result\maxlen}{.95\ht\strutbox}}%
  \makebox[\dimexpr\maxlen-\tabcolsep][r]{\toprint\%}%
}

\lstdefinestyle{codesnip}{
  basicstyle=\ttfamily\tiny,
  keywordstyle=\color{blue},
  commentstyle=\color{gray},
  numberstyle=\tiny\color{gray},
  numbersep=5pt,
  tabsize=2,
  breaklines=true,
  showspaces=false,
  showstringspaces=false,
  rulecolor=\color{black},
  moredelim=**[is][\color{red}]{@}{@}
}

\newcommand{\parax}[1]{\noindent\textbf{#1}}

\providecommand{\ie}{\textit{i.e.,}\xspace}
\providecommand{\eg}{\textit{e.g.,}\xspace}

\providecommand{\etal}{\textit{et al.}\xspace}

\providecommand{\wifi}{WiFi\xspace}

\newcommand{\ignore}[1]{}

\newcommand{\androzoodownloaded}{1,008,539\xspace}
\newcommand{\numbeaconsdks}{52\xspace}
\newcommand{\numbeaconapps}{9,976\xspace}
\newcommand{\numavailableinplaystore}{574K\xspace}

\usepackage{hyperref}
\usepackage{hyperxmp}

\reviewtrue

\setcopyright{popets}
\copyrightyear{YYYY}
\acmYear{YYYY}
\acmVolume{YYYY}
\acmNumber{X}
\acmDOI{XXXXXXX.XXXXXXX}
\acmISBN{}
\acmConference{Proceedings on Privacy Enhancing Technologies}
\begin{document}

\title[Your Signal, Their Data]{Your Signal, Their Data: An Empirical Privacy Analysis
of Wireless-scanning SDKs in Android}

\author{Aniketh Girish}
\affiliation{\institution{IMDEA Networks Institute / Universidad Carlos III de Madrid} \country{}}

\author{Joel Reardon}
\affiliation{\institution{University of Calgary / AppCensus} \country{}}

\author{Juan Tapiador}
\affiliation{\institution{Universidad Carlos III de Madrid} \country{}}

\author{Srdjan Matic}
\affiliation{\institution{IMDEA Software Institute} \country{}}

\author{Narseo Vallina-Rodriguez}
\affiliation{\institution{\mbox{IMDEA Networks Institute / AppCensus}} \country{}}

\renewcommand{\shortauthors}{Girish et al.}

\begin{abstract}

Mobile apps frequently use Bluetooth Low Energy (BLE) and WiFi scanning permissions to
discover nearby devices like peripherals and connect to WiFi Access Points (APs). 
However, wireless interfaces also serve as a covert proxy for geolocation
data, enabling continuous user tracking and profiling. This includes  
technologies like BLE beacons, which are BLE devices broadcasting
unique identifiers to determine devices' indoor physical locations; such beacons
are easily found in shopping centres. 
Despite the widespread use of wireless scanning APIs and their potential for privacy abuse, 
the interplay between commercial mobile SDKs with wireless sensing and beaconing
technologies remains largely unexplored. In this work, we conduct the first systematic
analysis of 52 wireless-scanning SDKs, revealing
their data collection practices and privacy risks. We develop a comprehensive
analysis pipeline that enables us to detect beacon scanning capabilities, inject
wireless events to trigger app behaviors, and monitor runtime execution on
instrumented devices. 
Our findings show that 86\% of apps integrating these SDKs collect at least one
sensitive data type, including device and user identifiers such as AAID, email,
along with GPS coordinates, WiFi and Bluetooth scan results.
We uncover widespread SDK-to-SDK data sharing and evidence of ID bridging, where persistent
and resettable identifiers are shared and synchronized within SDKs embedded in 
applications to potentially construct detailed mobility
profiles, compromising user anonymity and enabling long-term tracking. We provide
evidence of key actors engaging in these practices and conclude by proposing
mitigation strategies such as stronger SDK
sandboxing, stricter enforcement of platform policies, and improved transparency
mechanisms to limit unauthorized tracking.

\end{abstract}


\maketitle

\vspace{-1mm}
\section{Introduction} 
\label{sec:introduction}

Imagine walking through a shopping mall and, as you pass by stores, your
phone buzzes with personalized ads and notifications. This
seamless integration of the digital and physical worlds is powered by
technologies like GPS, Bluetooth, and \wifi, as their sensing capabilities 
facilitate detecting device proximity or location. However, the use of these
technologies raise significant privacy concerns because
geolocation data is \textit{inherently sensitive}: either coarse-
or fine-grained geolocation data can reveal
individuals' daily habits, work-home pairs~\cite{deMontjoye2013}, social
structures~\cite{heinrich2021can, yu2024security}, 
and visits to sensitive sites like places of worship~\cite{ftc2024xmode}. 
Unfortunately, the demand for location data by marketing, banking, and insurance firms has
fueled a complex supply chain of actors specializing in location data collection,
aggregation, and resale~\cite{tomkemp_data_brokers}. Companies like
Azira~\cite{azira2024} and Venntel~\cite{wsjVenntel2024}
have been reported selling location data, including visits to sensitive locations
such as abortion clinics~\cite{wyden2024} and even private sites like
Jeffrey Epstein's island~\cite{Epsteinwired2024}. 

Despite the permission mechanisms implemented in Android and iOS 
to protect access to precise GPS data~\cite{apple_location_services_privacy,
android_location_permissions}, app developers and third-party SDKs
exploit alternative channels to bypass these controls and obtain (or infer) users' location, 
often in the background without user awareness. 
One such mechanism is based on scanning wireless devices and beacon signals, 
including classic Bluetooth, Bluetooth Low
Energy (BLE) beacons, and \wifi access points.\footnote{In this paper, we
collectively refer to SDKs with BLE and \wifi scanning capabilities as ``beacon
SDKs''. Equally, for simplicity, we refer to BLE and \wifi broadcast messages as
``beacons.''}
Such wireless scans can reveal users' geolocation as
Bluetooth or \wifi beacons are tied to specific
physical locations like stores, hotels, workplaces, or transit hubs~\cite{cunche2012know}. 
Linking such geolocation data with user or device IDs like the Android Advertising ID (AAID)
and MAC addresses make reversing users' identity and movement patterns
straightforward~\cite{drakonakis2019please}.

The threats posed by the collection of wireless beacons as geolocation data
proxies are real and severe, as several high-profile incidents and legal actions show. 
In 2014, the Snowden revelations exposed how devices could be tracked via MAC
addresses through \wifi hotspots~\cite{snowden2014mac}. 
More recently, the Wall Street Journal reported that the FBI and ICE acquired location data through
SDKs embedded in mobile apps~\cite{wsj2023fbi, wsjVenntel2024}. 
The Federal Trade Commission (FTC) has taken action against companies like 
X-Mode (now Outlogic)~\cite{dehaye2020proximity},
inMobi, and Kochava for harvesting and reselling extensive location data without
user consent, leading to serious concerns about
constitutional rights in the U.S.~\cite{ftc2024xmode, inmobi_ftc,
 ftc2024location, ftckochava2024}. A recent report
by the German data protection authority reveals that apps collecting
location data, including military personnel movements, was sold in data
marketplaces, exposing intelligence agency sites, military bases, secret
facilities and their personnel's
habits~\cite{br2024,wired2024_us-soldiers-spies-nuclear-germany}. 
In the academic sphere, researchers demonstrated that wireless 
signal can be exploited to infer geographical locations 
and social structures~\cite{cunche2012know,rose2010mapping,matte2015device,
Gagnonrssi2024,greenstein2007can}. 
Reyes~\etal~\cite{reyes2018won} found evidence of 
children-directed apps using \wifi APs as a proxy
to GPS locations, violating the COPPA Rule.
Similarly, Girish~\etal~\cite{girish-imc23} identified apps and SDKs leveraging local network
scans to infer user location without user consent in smart home networks.

Despite these efforts, SDKs exploiting wireless scans for covert location
tracking remain poorly understood. Consequently, existing app auditing mechanisms and
privacy controls (including data randomization) have been proven insufficient to detect
and protect users against such
invasive practices, respectively~\cite{girish-imc23,vanhoef2016mac}. 
To fill this gap, we empirically analyze 
how \numbeaconsdks mobile beacon SDKs leverage GPS, BLE, and \wifi as side channels to 
covertly track individuals' proximity and locations, covering \numbeaconapps Android apps.
To uncover their scanning capabilities, data-sharing, and privacy risks,
we develop a hybrid analysis pipeline that 
combines (i) static analysis, including API usage detection and control-flow
analysis to study cross-library
interactions and (ii) dynamic analysis, using an instrumented
device that injects wireless events and signals at the OS-level to trigger SDK
behaviors and observe app runtime behaviors. To the best of our knowledge, 
this study is the first large-scale empirical
analysis of BLE and WiFi scanning SDKs in Android apps to systematically
uncover their characteristics, interrelationships, and privacy
implications. The key contributions of this work are:

\begin{itemize}[leftmargin=*]
\item We identify \numbeaconsdks SDKs with \wifi and BLE scanning features integrated into 
  at least \numbeaconapps apps with an estimated cumulative install count of 55B devices. 
  We find that beacon SDKs often offer a dual purpose, functioning both as  
  beacon enablers and as analytics or advertising libraries. Specifically,  
  43 SDKs support analytics, 40 SDKs provide location services, and 9 SDKs
  integrate advertising or user profiling features (\S\ref{subsec:prevalence}). Among them, we detect 28
  SDKs that play a key role in extensive cross-library interactions, potentially enabling
  colluding SDKs to silently share with each other data such as Bluetooth, \wifi
  scans, and geolocation for advertising and tracking purposes (\S\ref{subsec:cross_library_analysis}).

\item We observe that 86\% of beacon-enabled apps extensively collect personally
identifiable information (PII), along with GPS coordinates, \wifi, and Bluetooth
scan results (\S\ref{subsec:access_beacon_collection}). 
  However, their data
  collection practices extend further, with 19\% of beacon SDKs engaging in ID
  bridging by linking persistent IDs (e.g., Android ID and MAC addresses) 
  with resettable ones
  (e.g., AAID) to build detailed user mobility profiles,
  potentially violating Google Play Store policies (\S\ref{subsec:id_bridging}). These risks are exacerbated
  by the use of persistent proprietary IDs (e.g., Adobe's Marketing Cloud ID), 
  unvetted cross-library data sharing, and by exploiting shortcomings
  of Android's permission and sandboxing models.
  Moreover, 71\% of apps requesting location permissions fail to provide a  
  rationale through Android's \textit{shouldShowRequestPermissionsRationale()}
  API, leaving users unaware of why access is requested (\S\ref{subsec:consent_rationale}). We also observe
  SDKs abuse vulnerabilities in unpatched devices to bypass the 
  Android permissions governing BLE and \wifi permissions.

\item These widespread practices undermine existing privacy protections and  
  controls. In \S\ref{sec:discussion}, we propose mitigation strategies, including:  
  SDK sandboxing to restrict cross-library data sharing,  
  performing runtime audits to detect unauthorized tracking,  
  enforcing stricter platform policies to curb ID bridging, and  
  improving transparency mechanisms by requiring clear permission rationales  
  and explicit disclosures of SDK data access practices.  
  Additionally, we advocate for stronger regulatory oversight, platforms' 
  permission re-designs
  and stricter vetting processes (i.e., Google Play Protect)
  to proactively conduct large-scale audits  
  and enforcement actions against non-compliant tracking practices.

\end{itemize}

\parax{Responsible disclosure.} We shared a preprint of this paper with
Google as operator of the Android platform, and with the European Data Protection
Supervisor (EDPS), the Spanish Data Protection Agency (AEPD), and the
French National Commission on Informatics and Liberty (CNIL).

\parax{Research artifacts.} To promote transparency, reproducibility, and future
research, we will release the final dataset, beacon detection scripts, static
analysis pipeline, and analysis code used in this study, excluding proprietary
components. These artifacts will be publicly available at:
\url{https://github.com/wireless-scanning-SDKs/}.

\section{Background}
\label{sec:background}
 
This section provides an overview of BLE (\S\ref{background:ble}) and WiFi
(\S\ref{background:wifi}) beacon technologies as location proxies, and an overview of 
the existing Android permissions for restricting apps' access to BLE and
\wifi scanning capabilities (\S\ref{background:perms}).

\subsection{BLE Beacons}
\label{background:ble}

A BLE beacon is a non-pairing device broadcasting unique IDs like UUIDs or MAC addresses to infer
proximity or location. 
Unlike GPS-based positioning systems, BLE beacons are deployed in 
fixed locations---e.g., stores, metro stations or offices---so that they
provide fine-grained indoor location for 
like retail~\cite{itransition_beacons}, healthcare, logistics, 
or crowd management at events~\cite{neoma_crowd_management}. 
These fixed signals are collected alongside GPS coordinates and are often
aggregated into publicly accessible databases such as
Wigle~\cite{wigle_networks}, and proprietary ones.
This allows SDKs to infer a user's location solely through nearby BLE
signals and access to such datasets. 

BLE beacons are not a Bluetooth SIG standard; rather, they are custom
standards developed by large providers or groups of
companies. Popular BLE beacon solutions include iBeacon
by Apple~\cite{apple_ibeacon}, the now deprecated Eddystone by
Google~\cite{google_eddystone}, and the widely supported open-source
and platform-agnostic AltBeacon~\cite{radius_altbeacon}, which is
becoming a predominant solution due to their
cross-platform interoperability.
We note that standard Bluetooth devices can be also 
scanned to collect device names and MAC addresses, which could be used for 
 environmental sensing (\eg by inferring locations based
on paired BLE devices), 
and social network inference~\cite{lyons2023log}.

\subsection{\wifi Beacons}
\label{background:wifi}

\wifi Access Points (APs) broadcast beacon frames containing their MAC
addresses and received signal strength indicator (RSSI) over 802.11b/g/n channels. Mobile devices can passively
scan these beacons to locate and connect to nearby APs by sending association frames.
\wifi APs are typically deployed in known locations (\eg homes,
shops, hotels, and airports) and rarely move once
deployed. This means that correlating
\wifi SSID and BSSID data\footnote{SSID is the name of a wireless network that devices use to identify and connect to it. 
 BSSID is the unique identifier for a specific access point within a wireless network (i.e., MAC 
address).}
with free online databases (e.g., \url{wigle.com}) or commercial location services
(\eg \url{here.com}) makes them highly effective for determining user location.
In fact, there are commercially-available \wifi Positioning System (WPS) that
facilitate inferring users' locations from \wifi signals, as a complement
or alternative to GPS-based systems.
Apple integrates WPS~\cite{apple_location_services} into iOS devices for
enhanced location services, though Rye~\etal~\cite{rye2024surveilling}
recently highlighted the potential for mass surveillance using this
data. Similarly, Google maintains a WPS populated through \wifi AP
data collected from Android devices and its ecosystem, which also serves as
a data source for Android's coarse location
permission~\cite{google_geolocation_api,vallina2013assistance}. 
These WPSes expose APIs that
allow a client to submit \wifi scan data and receive location estimates in
return. Other providers, such as Skyhook
Wireless~\cite{skyhook_wifi_location} and Navizon~\cite{navizon}, offer
hybrid systems combining GPS, \wifi, and cellular signals to triangulate
device location.

\subsection{Permission Overview} 
\label{background:perms}

Android's permission model regulates access to \wifi and Bluetooth scanning
functions 
to protect user privacy and security~\cite{android_permissions_overview}.
Over successive releases, Android's permission model has evolved to impose
stricter controls on BLE and \wifi scanning capabilities, particularly due to their
dual usage as location proxies. 

\vspace{1mm}
\noindent \textbf{Bluetooth.\xspace}
Initially, Bluetooth operations required only a
generic permission for scanning
and connecting with nearby devices. 
However, starting
with Android 6, apps also had to request the runtime\footnote{Unlike normal
permissions, which are granted during installation, \textit{runtime} or \textit{dangerous}
permissions protect sensitive resources and data (\eg location) and
require explicit user approval at runtime~\cite{android_permissions_overview}.}
\path{ACCESS_COARSE_LOCATION} permission, as BLE scans could indirectly reveal
location data~\cite{android6change}. This was later upgraded in Android
10, mandating the \path{ACCESS_FINE_LOCATION} permission for Bluetooth scanning
due to its increased risk of exposing the precise location
data~\cite{karpiniec2023bluetooth_permissions}. 
However, this security improvement only led to
confusion among users, especially when apps that did not inherently need location
data---such as those for connecting to Bluetooth peripherals---were also forced to 
request location
permissions~\cite{wright_android12_bluetooth}. To address this issue, Android 12
introduced a finer-grained model with three new permissions: one for scanning
(\path{BLUETOOTH_SCAN}), one for establishing connections
(\path{BLUETOOTH_CONNECT}), and one to advertise as a peripheral
(\path{BLUETOOTH_ADVERTISE}). Developers must request both \path{ACCESS_FINE_LOCATION} and
\path{ACCESS_COARSE_LOCATION} to allow users to choose between precise or
approximate location access at runtime. Yet, without clear context, users
may unintentionally grant excessive location privileges—e.g., precise location
for Bluetooth pairing. To mitigate this, Android 12 introduced the
\texttt{neverForLocation} flag in 2021 to ensure that Bluetooth scan data is not
misused for location inference.  

\vspace{1mm}
\noindent \textbf{\wifi.\xspace}
Since Android's first release, apps using \wifi must request the
\path{CHANGE_WIFI_STATE} and \path{ACCESS_WIFI_STATE} permissions to toggle
connectivity, modify settings, and provide connection details.
Since Android 8, any app that scans the \wifi network is also required to
request at least one location permission
(i.e., \path{ACCESS_FINE_LOCATION} or \path{ACCESS_COARSE_LOCATION}).
More recently, Android 13 introduced a new runtime permission,
\path{NEARBY_WIFI_DEVICES} for connecting to nearby devices
via \wifi. Android 13 also added support for the \texttt{neverForLocation} tag
to specify \wifi scan data is not used for location inference.

\vspace{1mm}
\noindent \textbf{Platform Fragmentation.\xspace}
Due to the high version fragmentation in the Android ecosystem, many apps and
SDKs intentionally exploit OS vulnerabilities and old (typically less restrictive) versions of
the Android permission model to perform BLE and \wifi scans. 
Another key limitation of the permission model is its inability to
separate between permissions requested by the host app and those by
third-party SDKs~\cite{feal2021don}, 
even for runtime permissions that grant access to sensitive data.
As a result, SDKs can piggyback on the permissions requested by the host 
app for secondary purposes such as surveillance or advertising.

\begin{figure}[t!]
    \centering
    \small
    \includegraphics[width=0.90\columnwidth]{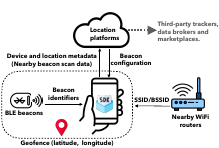}
    \caption{Beacon data ecosystem.}
    \label{fig:beacon_ecosystem}
\end{figure}

\section{Threat Model}
\label{sec:threat_model}

We consider an adversary whose primary objective is to
collect geolocation data from mobile Android apps  
by scanning nearby devices using Bluetooth or
\wifi permissions (see Figure~\ref{fig:beacon_ecosystem}).
This adversary can be any party with programmatic access to the device for 
scanning data, including the app developer (the first party) 
as well as third party SDKs
offering advertising, analytics, or location-based solutions. We consider
three key threats posed by such adversaries:

\begin{enumerate}[leftmargin=*]
\item \parax{Continuous Location Tracking through Proxy Sensors.}
Wireless-enabled location SDKs or Android apps can \textit{silently} track users
by continuously scanning for nearby \wifi and BLE signals. Scan results can
include: device information (e.g., device name), network information (e.g.,
SSID/BSSID, MAC addresses), and other unique device IDs as in the case of BLE
beacons. To infer user location, wireless scanning data can be correlated with
external databases that map MAC addresses, beacons and \wifi AP BSSIDs and SSIDs
to geographic coordinates as described in the previous section. SDKs often
enrich scan data with precise GPS coordinates and other device or user
identifiers like the AAID to construct detailed mobility profiles of millions of
users. ~\footnote{The Android OS allows app and SDK developers to
programmatically access a wide range of IDs with varying
properties~\cite{android_user_data_ids}. The MAC address of a \wifi
AP are device-specific, persistent and globally unique IDs. Similarly, the
user's email address, even if hashed, is a user-specific, persistent and unique
ID, unless it is shared by a group of users. The AAID is a device-specific,
resettable and globally unique ID\@. The Firebase ID, instead, is a globally
unique and app-specific ID that is resettable by reinstalling the app to which
it is scoped.} Table~\ref{tab:pii_table} in the Appendix provides an exhaustive
description of the PII types and user/device IDs considered in this paper,
categorized by their nature. These rich user profiles may reveal sensitive
information such as social structures, frequently visited locations and
lifestyle patterns~\cite{drakonakis2019please}. 
  
\item \parax{ID Bridging.} Beacon information facilitates collecting unique user
geolocation fingerprints---as demonstrated by de
Montjoye~\cite{deMontjoye2013}---due to the uniqueness of human mobility
patterns. Mobility fingerprints not only allow distinguishing individuals even
in anonymized datasets but also enable ID bridging, i.e., correlating and bridging
users data and fingerprints for continuous user
tracking and profiling. These methods, particularly when used to bridge
resettable IDs like the AAID, significantly undermine user attempts to
maintain anonymity on the Internet, even if users opt-out of personalized ads
through system settings or by resetting their AAID. To mitigate these privacy risks,
Google has set various policies and best practices to inform and educate app
developers about which IDs would be better for specific
purposes~\cite{googleplay_user_data_policy}. Google strictly prohibits linking
resettable IDs (\eg AAID) with persistent or global IDs for advertising or
analytics purposes without sufficient
transparency~\cite{googleplay_user_data_policy, android_user_data_ids}. 
\S\ref{subsec:id_bridging} analyzes ID bridging practices across
beacon SDKs, including AAID bridging, while \S\ref{subsec:cross_library_analysis}
shows how colluding SDKs programmatically interact with each other 
to synchronize user profiles without user awareness. 
Oftentimes, geolocation data is aggregated and sold to 
advertisers, analytics companies, and other entities like data brokers, 
causing significant privacy harm to users.

\item \parax{Exploitation of Outdated Permission Models.} SDKs can exploit
  vulnerabilities in Android's permission model (\S\ref{background:perms})
  to access sensitive data in unpatched devices.
  Methods to achieve this include piggybacking on permissions
  granted to the host app, using legacy APIs with less stringent requirements,
  or leveraging side channels to bypass restrictions and infer information
  without explicit permission requests.
  This type of attack is facilitated by Android's version fragmentation:
  according to Statista, over 41\% of Android devices run version
  12 or below as of January 2025.
  \S\ref{subsec:access_beacon_collection} reports evidence 
  of SDKs using these techniques to access location data.

\end{enumerate}

\begin{figure}[t!]
    \centering
    \includegraphics[width=\columnwidth]{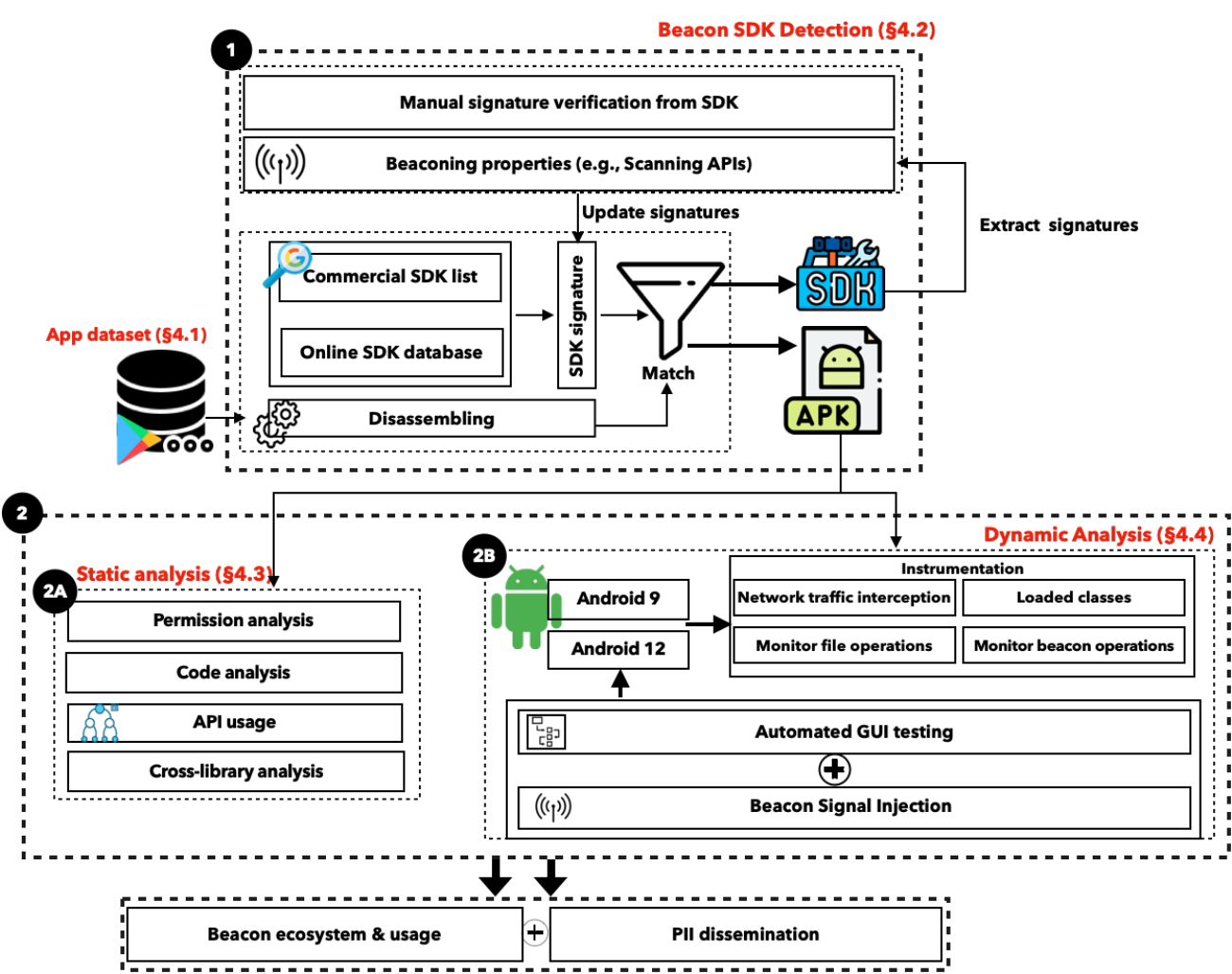}
    \caption{Methodology overview. Processes 2.a and 2.b. run in parallel.}
    \label{fig:methodology_overview}
\end{figure}

\section{Methodology} 
\label{sec:methodology}

Figure~\ref{fig:methodology_overview} summarizes the methodology we developed
to achieve the paper objectives.
We combine both static and dynamic analysis techniques to 
comprehensively detect and analyze how apps collect geolocation data 
through BLE and \wifi beacons, its dissemination with user IDs, 
and the SDKs offering such solutions. We
note that, due to the inherent limitations of black-box testing, we
do not claim completeness. However, our methods provide actual evidence of privacy
concerns associated with the use of wireless scanning capabilities by SDKs. 
We next describe each of the components of our methodology.

\subsection{Dataset}
\label{subsec:dataset}

We use the AndroZoo dataset~\cite{androzooallix2016} to collect
\androzoodownloaded apps uploaded after January 2023. AndroZoo
is a large-scale collection of Android apps that includes over 10 million
items historically indexed on the Google Play Store. 
We filter this dataset to select \numavailableinplaystore
apps available in the Google Play Store when accessed from the
European Union (EU) as of July 2024, ensuring that our study covers the
most recent app versions. 
We use an open-source Google Play scraper~\cite{Olanoscraper2024} to
collect metadata for each app, including its self-disclosed app 
category, developer information, data
safety labels, and install counts.

\subsection{Detecting Beacon SDKs}
\label{subsec:beacon_sdk_detection}

There is no publicly available inventory of SDKs offering BLE and WiFi scanning
capabilities. While some efforts document advertising and analytics 
SDKs~\cite{malibradar2016,razaghpanah2018apps,exodusprivacy2024,feal2021don,libid},
beacon-enabled SDKs remain under-reported, often embedded within broader
functionalities, making them difficult to detect and analyze. To address this gap, we create
a first-of-its-kind public dataset, capturing a representative set of widely
used SDKs with wireless scanning implementations. Our approach systematically identifies
these SDKs through a multi-step iterative process, combining third-party
datasets and manual searches to curate a structured list of SDKs actively
engaging in wireless scanning.

\begin{itemize}[leftmargin=*]
  
\item \parax{Step 1: Identifying Known SDKs.} To identify SDKs that perform BLE
and WiFi scanning, we begin with an initial set of SDKs listed by
Exodus~\cite{exodusprivacy2024} and Libradar~\cite{malibradar2016}, two widely
used SDK detection tools. %
While these tools cover popular advertising and analytics SDKs and label them based on their capabilities
(including BLE and WiFi capabilities), they are not
comprehensive, potentially missing newer or less-documented SDKs as well
as scanning-specific SDKs. From these
sources, we identify an initial list of 14 SDKs related to BLE/WiFi scanning.

\item \parax{Step 2: Targeted Online Searches.} We expand the initial list of SDKs 
identified in Step 1 
by conducting web searches for commercial SDKs offering proximity, location, and
Bluetooth/WiFi scanning functionalities. For that, we perform targeted keyword searches using
Google's search engine.~\footnote{We use keywords such as ``beacon SDK,''
``proximity SDK,'' ``BLE SDK,'' and ``\wifi beacon SDK\@.''}
This process identified 22 additional SDKs that were not included in the initial
list. However, some results are false positives triggered by keyword matches
while the SDKs do not implement beacon scanning. Hence to confirm functionality, we
manually review official documentation, API references, integration guides, and
discussions in SDKs' developer forum, along with source code repositories (where
available) to identify SDKs that explicitly invoke BLE or WiFi scanning
capabilities.~\footnote{We manually identified more than 200 system APIs
related to BLE and \wifi scanning from Android's official
API reference and AOSP (Android Open Source Project) source code. These methods are released
as part of our open-science approach.}

\item \parax{Step 3: Refining \& Expanding SDK Selection Through Static
Analysis.} We refine and expand the set of SDKs populated in the
previous steps, by applying the static analysis techniques illustrated in
\S\ref{subsec:static_analysis} to automatically analyze a subset of 362K apps invoking methods indicative of beacon scanning.
These features are extracted from the
curated list of SDKs identified in Steps 1 and 2, producing
\textit{SDK signatures} for each SDK that cover from system API calls related
to BLE and WiFi scanning (e.g.,
`WifiManager.startScan()'~\cite{androidbluetooth2024,androidwifiscan2024})
to SDK attribution signals such as class names,
URL endpoints, and custom permissions. Additionally, we search for special
functionalities such as UUID broadcasting and RSSI analysis, which further
indicate BLE/WiFi scanning behavior. For reference, an example set of signals
is provided in Table~\ref{tab:beacon_apis} in the Appendix. After applying our
refinement process, we identified a total of 55 SDKs. 

\end{itemize}

\parax{Validation.} 
To ensure high-confidence detection and minimize false positives, we manually
analyze %
the code of up to three representative apps per SDK to confirm
whether these SDKs implement scanning capabilities. This check allowed us to
remove three SDKs designed for general connectivity with peripherals
rather than for wireless scanning. At the end of this manual process, we
curate a final signature set of \numbeaconsdks beacon SDKs features and signals,
which we use to statically identify at scale apps integrating these SDKs.

\subsection{Static Analysis}
\label{subsec:static_analysis}

\hyphenation{should-Show-Request-Per-miss-ion-Ration-ale}

We apply static analysis to examine how apps integrate and interact with
identified beacon SDKs. Specifically, we look for unique SDK signatures
(\S\ref{subsec:beacon_sdk_detection}) to identify their presence across apps,
including app manifest metadata (declared permissions, services, and intent
filters) and code-level attributes. We also inspect
execution paths via control flow analysis 
to determine how apps invoke, interact with, or delegate
scanning capabilities to other integrated SDKs, as we describe next. Our static
analysis follows accepted research practices, focusing on studying public code structures
and API calls, ensuring that no proprietary insights or trade secrets are
extracted~\cite{GambaSP2020,arzt2014flowdroid,wang2021understanding, poeplau2014execute}.

\parax{Permission Analysis.} We use a custom-built parser to analyze
the manifest of each app, extracting AOSP and custom permissions requested by apps
and beacon SDKs, along with their declared services, providers, and receivers. We
also analyze metadata such as the \texttt{neverForLocation} attribute which
indicate whether location permissions are used exclusively for non-location purposes, 
and the \texttt{shouldShowRequestPermissionRationale()} API,
which provides an educational UI explaining why a specific permission is needed
to enable a feature. App developers should declare these attributes to ensure
transparency so their absence can indicate a lack of proper disclosure. These
elements are studied in \S\ref{subsec:consent_rationale} to assess
beacon SDKs' transparency over data collection.

\parax{Code Analysis.} 
We use Androguard~\cite{androguard2024} to extract Dalvik bytecode and identify
SDK signatures (\eg class names and APIs) that indicate beacon SDK functionality
and data dissemination. These signatures are detailed in
Table~\ref{tab:sdk_signatures}.
To analyze execution paths, we construct control flow graphs (CFGs) for each
app, where nodes represent methods or functions (with associated class names),
and edges denote function calls between them. CFGs trace execution from source
methods (e.g., Android lifecycle methods like onCreate() and onResume(), where
SDKs typically initialize) to sink APIs that collect sensitive data, such as
location (e.g., LocationManager.getLastKnownLocation()) or beacon data (e.g.,
WiFi scan results). A mapping of API calls considered
in this process is provided in Table~\ref{tab:system_apis}.
To attribute SDKs and identify SDKs actively being used rather than dead code, 
we match class names in CFG nodes against known
SDK signatures extracted from decompiled app code. 

\parax{API Usage.} To analyze how beacon-enabled apps interact with
sensitive resources, we traverse the CFG to trace data
flows.  Starting from sink APIs (e.g.,
`LocationManager.-getLastKnownLocation()'), we visit nodes backward to locate the
source methods invoking these APIs. This approach links
sensitive API calls to the Android permissions protecting them, allowing us
to study the data flows within the app and also third-party SDKs that may
piggyback on these permissions (\S\ref{sec:permission_analysis}). As an
official and updated comprehensive permission mapping database is not available, we
construct one by combining: (1)
pre-existing permission mappings from PScout~\cite{Au2012PScout},
Androguard, and Axplorer~\cite{Backes2016Axplorer}; and (2) mappings
derived from the \texttt{@RequiresPermission} annotations in the AOSP source code,
which we publicly release as part of our research artifacts.

\parax{Cross-Library Analysis.} We statically investigate interactions between
SDKs embedded within the same app to identify dependencies and data-sharing practices
between co-located SDKs. Specifically,
we identify cross-library interactions by studying function calls where one SDK code
invokes methods or accesses data from another SDK\@. We attribute these
interactions among SDKs by detecting function calls between classes associated
with beacon SDKs. To distinguish cross-library interactions from app-internal
calls, we compare the class names of the caller and callee classes: if their
top- and second-level domains (1) do not match each other, and (2) do not match
the host app's package name, the interaction is flagged as cross-library. 
We manually validate all detected interactions to avoid mis-reporting cross-SDK
interactions.

\vspace{-1mm}
\subsection{Dynamic Analysis}
\label{subsec:dynamic_analysis}

Dynamic analysis allows us to gather actual evidence of beacon SDK runtime
behavior by executing apps on instrumented devices~\cite{reyes2018won, achara2014short, kollnig2021iphones, reardon201950}. 
Specifically, we
observe how apps interact with sensitive resources, handle beacon data
(e.g., BLE, \wifi), transmit network traffic, and respond to injected BLE and WiFi
signals. To achieve this, we use a device farm comprising eight Android 9 and
eight
Android 12 Google Pixel 3a smartphones located in the EU, each running an
instrumented version of the operating system. 
This dual-version approach allows us to analyze how app behaviors adapt to
changes in the Android permission model across different versions and
how SDKs may abuse legacy versions.

\parax{Instrumentation.} We use a customized AOSP version to transparently
monitor runtime resource access, such as accesses to permission-protected
APIs, BLE/\wifi scans, I/O file operations, and capturing all network
traffic, by instrumenting the relevant methods and APIs to capture
their activity.
We observe reads and writes to TLS sockets by instrumenting relevant APIs,
allowing us to analyze network traffic without interfering with the TLS
handshake or using our own certificates. We also decode common encodings like
gzip and base64, along with bespoke obfuscation methods 
used by popular SDKs (e.g., Forter, JPush, and Yandex)
to identify disseminated sensitive data,
as discussed in \S\ref{subsec:access_beacon_collection}.
To complement our static
analysis-based SDK detection (\S\ref{subsec:static_analysis}), we
instrument the Android Runtime (ART) to log classes loaded at runtime by
tracking the \texttt{FindClass} method of the class linker to detect
classes loaded during execution. This approach effectively increases the
reliability of static signals and enables the identification of obfuscated
classes that static analysis might miss. 

\parax{Automatization.} We use Android Monkey~\cite{AndroidMonkeyTool} to automate app execution with 
synthetic UI inputs for 8-10 minutes. To bypass registration walls,
each device is configured with unique pseudonymous identifiers (e.g., phone
numbers, email addresses, and usernames), allowing us to trace
the potential dissemination of this data in network flows for ID bridging. 
Our automation also handles standard
logins, including Single Sign-On (SSO) flows like Google SSO whenever
applicable.

\parax{Beacon Signal Injection.} In addition of automatizing UI inputs, 
it is essential to simulate realistic beacon signals to trigger runtime responses in SDKs implementing
them. 
We achieve this by injecting Bluetooth and \wifi scan results during
app scans, simulating
nearby devices or networks directly on our AOSP instrumentation. 
The injected BLE advertising data
conforms to standard beacon formats and includes beacon types such as iBeacon,
AltBeacon, Eddystone (URL, UID), and GAEN beacons, formatted according to their
respective standards~\cite{apple_ibeacon, radius_altbeacon, google_eddystone,
apple_gaen, google_gaen}. Additionally, we use a Raspberry Pi 4 to replay intercepted 
BLE beacon payloads and MAC addresses every 5 seconds, with the goal of exercising further app
activity upon beacons detection. 
Spurious \wifi networks are also injected, including
SSIDs and BSSIDs, as well as SSIDs appended with
\texttt{\_nomap} and \texttt{\_optout}, to evaluate whether SDKs respect
router-owners preferences to exclude their devices from scanning. 
We use distinctive palindromic MAC addresses in the
injected results to monitor their dissemination in TLS flows
to the cloud. 

\subsection{Limitations}
\label{subsec:limitations}

As with any empirical measurement study, our analysis is inherently constrained
by the black-box nature of testing methods, the opacity of SDKs, and the continuous
evolution of implementations. In fact, our list of beacon SDKs
extracted with static analysis methods may not offer a complete
picture of this complex landscape due to code obfuscation, reflection, and
dynamic loading~\cite{continella2017obfuscation, faruki2014evaluation,
maiorca2015stealth}. 

Dynamic analysis successfully executes 97\% of the apps, despite challenges
like root detection, certificate pinning, emulator checks, and behavioral
fingerprinting, which may limit visibility into SDK behaviors. Beacon injection
effectively triggers scanning in most cases but may not trigger SDKs relying on
custom or proprietary formats that are not publicly documented. Additionally,
Android Monkey automates standard interactions, though it cannot bypass CAPTCHA
or 2FA logins yet prior work~\cite{reyes2018won} shows it explores code paths
60\% similar to human interaction. 

We run our experiments on EU-based devices where
stricter privacy rules apply so it is possible that
specific SDK behaviors may vary in jurisdictions with more permissive regulations. 
Nevertheless, these limitations do not compromise the validity of our findings:
while we do not claim completeness, our methodology effectively uncovers
previously undocumented tracking behaviors within beacon SDKs, highlighting
systemic privacy risks and the need to regulate their usage.

\vspace{-1mm}
\section{Beacon SDK Landscape}
\label{sec:ecosystem}

\begin{table}[t!]
  \centering
  \LARGE
  \caption{Top 20 SDKs in the dataset, classified depending on their beacon types:
    Bluetooth, WiFi or they operate as integration partners. For the purpose
    of each SDK, in addition to analytics and location, we mark SDKs that
    offer advertising ($\bullet$) and profiling ($\diamond$) services.}
  \resizebox{\columnwidth}{!}{%
    \begin{tabular}{lrcccccr}
      \toprule
      \textbf{\makecell{SDK Name}} & \textbf{\# Apps} & \textbf{\makecell{Total \\ Installs}} & \multicolumn{3}{c}{\textbf{\makecell{Beacon \\ Type}}} & \multicolumn{2}{c}{\textbf{\makecell{Purpose \\ Type}}}  \\
      \cmidrule(lr){4-6} \cmidrule(lr){7-8}
      &  & & {\textbf{BLE}} & {\textbf{WiFi}} & {\textbf{Integration}} & {\textbf{Analytics}} & {\textbf{Location}}  \\
      \midrule
      AltBeacon                            & 4,022 & 5B    & \checkmark &            &            & \checkmark & \checkmark \\
      Adobe Experience Platform                       & 1,328 & 8B    &            &            & \checkmark & \checkmark &            \\
      Kochava~$\bullet$                    & 1,117 & 15B   &            & \checkmark & \checkmark & \checkmark & \checkmark \\
      Salesforce Marketing Cloud~$\bullet$ & 1,080 & 6B    &            &            & \checkmark & \checkmark & \checkmark \\
      Estimote                             & 510   & 201M  & \checkmark &            &            & \checkmark & \checkmark \\
      LeanPlum~$\diamond$                  & 456   & 8B    & \checkmark & \checkmark &            & \checkmark & \checkmark \\
      Gimbal~$\bullet$                     & 396   & 3308M & \checkmark & \checkmark &            & \checkmark & \checkmark \\
      Radius Networks                      & 369   & 359M  & \checkmark &            &            & \checkmark &            \\
      mParticle                            & 367   & 2B    &            &            & \checkmark & \checkmark &            \\
      Ad4Screen~$\bullet$                  & 198   & 1B    & \checkmark & \checkmark &            &            &            \\
      Kontakt                              & 195   & 31M   & \checkmark &            &            & \checkmark &            \\
      CueAudio                             & 190   & 9M    &            &            &            & \checkmark & \checkmark \\
      Swrve~$\diamond$                     & 153   & 2B    & \checkmark &            &            & \checkmark & \checkmark \\
      Reveal Mobile                        & 109   & 6M    & \checkmark & \checkmark &            & \checkmark & \checkmark \\
      Exponea                              & 99    & 191M  &            & \checkmark & \checkmark & \checkmark &            \\
      Radar                                & 93    & 482M  & \checkmark &            &            & \checkmark & \checkmark \\
      IndoorAtlas                          & 92    & 7M    & \checkmark & \checkmark &            & \checkmark & \checkmark \\
      SignalFrame                          & 89    & 56M   & \checkmark & \checkmark &            &            & \checkmark \\
      Bazaarvoice                          & 88    & 420M  &            &            & \checkmark & \checkmark & \checkmark \\
      Huq Sourcekit                        & 81    & 347M  &            & \checkmark & \checkmark & \checkmark & \checkmark \\
      \bottomrule
    \end{tabular}
  }
  \label{table:trackers}
\end{table}

This section studies the beacon SDK landscape, %
focusing on their purpose,
market share (\S\ref{subsec:prevalence}), and integration capabilities to
leverage complementary features and data sharing (\S\ref{subsec:cross_library_analysis}).

\subsection{SDK Purposes and Prevalence}
\label{subsec:prevalence}

Using our SDK detection method (\S\ref{subsec:beacon_sdk_detection}), we find
the 52 SDKs with beacon capabilities embedded in at least 9,976 apps
collectively installed on 55B devices.\footnote{Google Play install counts
provide an approximate range of market reach, not exact adoption
figures. Prior work has used them as a standard reference for estimating user
base and impact, though they have limitations such as the exclusion of
pre-installed or sideloaded apps~\cite{beyond_gplay2018}.} 
The market share and feature set of beacon SDKs varies significantly.
Table~\ref{table:trackers} lists the top-20 SDKs categorized by beacon type
(BLE, WiFi, or integration partners) and their primary and secondary purposes
(\eg analytics, location services, or advertising). 
A full list is provided in Table~\ref{table:full_tracker_list} in the Appendix.
The most widely used SDKs are AltBeacon (40\% of apps), Adobe Experience Platform 
 (13\% of apps), Kochava (11\% of apps), 
Salesforce Marketing Cloud (11\% of apps) and Estimote (5\% of apps),\footnote{These
percentages are relative to the total number of
applications found with at least one beacon SDK ($N=9,976$).} 
which are collectively embedded in apps with a total of 37 billion installs. 
While we find beacon SDKs in apps from
every Google Play store categories, they are more prevalent in Lifestyle (12\%), Shopping
(9\%), and Sports (8\%), with BLE and \wifi beacon SDKs being
more prominent in specific market sectors:

\begin{itemize}[leftmargin=*] 
  \item BLE beacon SDKs are found in 54\% of the apps with a total install count of 55B
  devices. They dominate the Lifestyle (20\%) and Business (12\%) app categories, 
  with SDKs
  like Radius Networks (44\%) and Estimote (34\%) more likely
  integrated in Lifestyle apps. 
  Interestingly, all BLE beacon SDKs also support geofencing for
  location-based services, including targeted advertising and proximity
  marketing (\eg Radius Networks and 
  Estimote~\cite{radius_networks_proximity_kit,estimote_proximity_sdk}).

  \item WiFi beacon SDKs are found in 17\% of the apps, totaling 72B installs. 
  They are commonly found in Sports (16\%) and News \& Magazines (10\%) apps.
  The SDKs Yinzcam Sobek and Gimbal SDKs~\cite{gimbaladplatform}
  specialize in sporting events and live stadium experiences.
  Anecdotally, some apps in these categories also embed audio beacon SDKs (\eg
  CueAudio) to synchronize location data with contextual audio signals,
  enhancing real-time interaction and audience tracking~\cite{cue_unforgettable_experiences}.
\end{itemize}

\parax{Capabilities.}
Some of the beacon SDKs identified like Kochava
or Adobe Experience Platform are well-known advertising and tracking services~\cite{razaghpanah2018apps}. 
We analyze the capabilities offered by
beacon SDKs by inspecting their public documentation (see
\S\ref{subsec:beacon_sdk_detection}).
We observe that most of them claim to offer multiple features to app developers, primarily analytics
services (43 SDKs, with a total install count of 52B installs), 
location services (40 SDKs, 41B installs), advertising (9 SDKs, 21B installs), 
and user profiling (5 SDKs, 11B installs). These 
features are not mutually exclusive. For example, while the
the main business area of Kochava, Salesforce Marketing Cloud, 
and Adobe Experience Platform is an advertising, analytics and identity graph
services, respectively, they also integrate beacon scanning capabilities in 
some of their SDK versions to enhance their analytics and advertising businesses with
location data. This increases the privacy risks for consumers.

\subsection{Cross-Library Analysis}
\label{subsec:cross_library_analysis}

\begin{table}[t!]
    \centering
    \SMALL
    \caption{Top 10 most common SDK combinations. Count is the number of apps embedding each combination.}
    \label{tab:sdk_combinations}
    \resizebox{0.8\columnwidth}{!}{%
    \begin{tabular}{lc}
      \toprule
                             \textbf{SDK Combination} &  \textbf{Count} \\
      \midrule
                                 (AltBeacon, Kontakt) &    270 \\
                         (AltBeacon, Radius Networks) &    228 \\
                                   (CueAudio, Gimbal) &    173 \\
      (AltBeacon, Areametrics, Cuebiq, Reveal Mobile) &    117 \\
                  (Adobe Experience Platform, Gimbal) &    100 \\
                                  (AltBeacon, Cuebiq) &     80 \\
                                (LeanPlum, mParticle) &     78 \\
              (AltBeacon, Salesforce Marketing Cloud) &     73 \\
                                (AltBeacon, Estimote) &     67 \\
                     (AltBeacon, Cuebiq, SignalFrame) &     59 \\
    \midrule
    Total Unique SDK Combinations                     &                  281 \\    
    Total Installs                                    &                  6B \\
    \bottomrule
    \end{tabular}
    } 

    \end{table}

\begin{figure}[!t]
    \centering
    \includegraphics[width=\columnwidth]{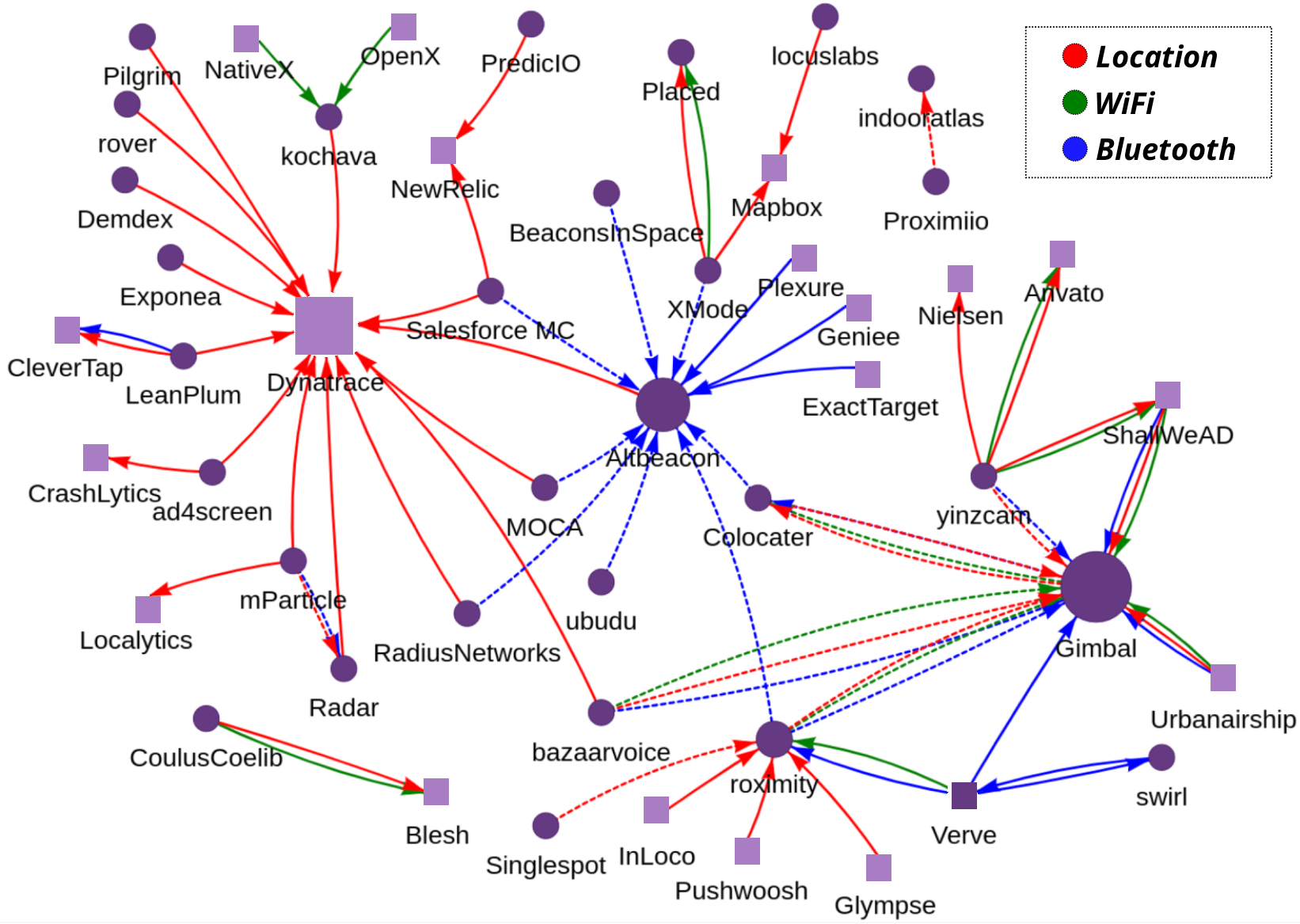}
    \caption{ Cross-library interactions between beacon and non-beacon SDKs (square nodes). 
	Dotted lines represent interactions between beacon SDKs, while solid lines 
	show interactions with non-beacon SDKs. Arrows indicate directionality, 
	from caller to callee SDK, with colors denoting the specific APIs accessed.}
    \label{fig:xlib_interactions}
\end{figure}

The 9,976 analyzed apps also include 331 non-beacon SDKs with further
data collection capabilities, including advertising and tracking services. 
We study the interactions between beacon SDKs with other 
SDKs (including non-beacon SDKs) that are
co-located within the same app using 
the methodology outlined in \S\ref{subsec:static_analysis}. 
Such co-existence allows SDK operators to exchange sensitive
data or complement their functionality including their tracking capabilities. 
In total, we identify 281 unique SDK combinations across the 9,976 apps containing at least one beacon
SDK. Table~\ref{tab:sdk_combinations} reports examples of such combinations, 
being AltBeacon and Kontakt the most common combination, found in 270 apps (195M users).

Figure~\ref{fig:xlib_interactions} captures SDK cross-invocations and data flows.
We detect 28 beacon SDKs exposing APIs
for cross-library interaction and data sharing to others. 
 We note that location APIs (red) 
are the most frequently invoked,
often transmitting data from beacon SDKs to advertising and analytics services.
WiFi-based APIs (green) are used for network-based geolocation, while BLE
APIs (blue) enable proximity-based functionalities. We classify
these interactions in two major categories:

\parax{Beacon SDK $\leftrightarrow$ Beacon SDK Interactions.}
We find 17 beacon SDKs
invoking APIs from one another. Notably, AltBeacon and Gimbal provide
APIs frequently used by other beacon SDKs to perform BLE scans and
access location data. For example, 
X-Mode, a location data aggregator, uses AltBeacon's APIs for
BLE scan results, while mParticle, a data integration platform, calls
Radar's APIs for location tracking. In some cases, multiple beacon SDKs
co-exist within the same app as in the French versions of McDonald's
and Burger King apps to locate user at their premises, 
both using AltBeacon for beacon scanning and indoor location
tracking, and Woosmap for geofencing. Woosmap also invokes APIs
from third-party SDKs, such as Urban Airship, Salesforce Marketing Cloud, and
Batch, via REST/OAuth APIs, enabling seamless data sharing and SDK-to-SDK
integration.

\parax{Beacon SDK $\leftrightarrow$ ATS Interaction.}
We identify 24 beacon SDKs sharing
BLE/WiFi scan results and geolocation data with 21 non-beacon
ATS SDKs. This behavior is concerning as advertising SDKs like
CleverTap and Twitter MoPub can use this
information for secondary purposes. Specifically, 11 ATS SDKs (\eg
UrbanAirship) invoke WiFi scanning APIs of 6 different beacon SDKs (\eg
Gimbal), and 16 ATSes invoke geolocation APIs from 22 beacon SDKs.
Interestingly, the Yinzcam Sobek SDK calls APIs from Fluzo, an Automatic Content
Recognition (ACR) SDK, for audio fingerprinting in the La Liga app (version
7.2.1), a potential GDPR violation that was investigated by the Spanish Data Protection
Agency (AEPD)~\cite{engadget2018spanish}. Such practices highlight the
complexity of data sharing and synchronization across-SDKs for secondary
purposes like advertising.

Unfortunately, these capabilities introduce unexpected security and privacy
concerns due to Android's coarse-grained sandbox and permission model. Since
there is no security boundary between libraries within the same app process, one
library can interact with another (\eg invoking functions) to access IDs and
sensitive data without restrictions or user
awareness~\cite{wang2021understanding}. Unlike the same-origin policy for limiting 
cross-site tracking on the web, Android's lack
of cross-SDK isolation mechanism allows them to access and
share data within the same app process without any restrictions.

\vspace{-1mm}
\section{Privacy Analysis}
\label{sec:sdk_dissemination}

Most of the analyzed apps with beacon SDKs (77\%) self-disclose the collection of user data through
Google Play's data safety labels. The most frequently collected data are
device IDs (62\%), personal information such as names, emails,
and other user IDs (60\%), and location (35\%) supposedly for app functionality (69\%) and
analytics (68\%), followed by account management (51\%), and advertising or
marketing (36\%). However, this information is self-disclosed by app developers
and may be erroneous or intentionally deceptive.

In this section, we use our dynamic analysis pipeline
(\S\ref{subsec:dynamic_analysis}) to analyze the runtime data collection practices
associated with beacon SDKs across our 9,652 apps.
Specifically, we study their permission 
requests (\S\ref{sec:permission_analysis}) and monitor
the dissemination of 18 sensitive data types including beacon-specific
data and user IDs (\S\ref{subsec:access_beacon_collection}), demonstrating how
beacon data is often bridged with device IDs like the
AAID (\S\ref{subsec:id_bridging}). To contextualize our permission
analysis in \S\ref{sec:permission_analysis}, we conclude with an
evaluation of how developers follow permission 
consent best practices (\S\ref{subsec:consent_rationale}). The
results presented in this section constitute actual evidence of the
dissemination of geolocation data linked to user IDs
from apps to cloud services and across SDKs,
 for potential secondary purposes like identity
profiling, mobility tracking, advertising and potentially data brokerage.

\subsection{Permission Analysis}
\label{sec:permission_analysis}

\begin{table}[t!]
    \centering
    \LARGE
    \setlength{\tabcolsep}{2pt}  %
    \caption{Usage of BLE, Location, and WiFi permissions by the top 15 SDKs, showing manifest declarations, API calls, and specific BLE\_SCAN tags introduced in Android 12+.}
    \resizebox{\columnwidth}{!}{%
    \begin{tabular}{lcccccccccccc}

    \toprule
    \textbf{SDK Name} & \textbf{\# apps} & \textbf{\shortstack{\# Android 12+}} & \multicolumn{2}{c}{\textbf{BLE}} & \multicolumn{2}{c}{\textbf{Location}} & \multicolumn{2}{c}{\textbf{WiFi}} & \multicolumn{2}{c}{\textbf{BLE\_SCAN}} \\
    \cmidrule(lr){4-5} \cmidrule(lr){6-7} \cmidrule(lr){8-9} \cmidrule(lr){10-11} \cmidrule(lr){12-13}
    &  &  & \textbf{Man.} & \textbf{API} & \textbf{Man.} & \textbf{API} & \textbf{Man.} & \textbf{API} & \textbf{\% Req} & \textbf{\% Tag} \\
    \midrule
    AltBeacon & 4024 & 1822 & \databarfull{95} & \databarfull{91} & \databarfull{96} & \databarfull{0} & \databarfull{73} & \databarfull{0} & \databarfull{57} & \databarfull{18} \\
    Adobe & 1328 & 514 & \databarfull{35} & \databarfull{0} & \databarfull{80} & \databarfull{0} & \databarfull{89} & \databarfull{0} & \databarfull{9} & \databarfull{25} \\
    Kochava & 1118 & 499 & \databarfull{20} & \databarfull{60} & \databarfull{42} & \databarfull{62} & \databarfull{97} & \databarfull{66} & \databarfull{5} & \databarfull{31} \\
    Salesforce & 1080 & 713 & \databarfull{33} & \databarfull{0} & \databarfull{80} & \databarfull{0} & \databarfull{72} & \databarfull{0} & \databarfull{12} & \databarfull{25} \\
    Estimote & 510 & 197 & \databarfull{97} & \databarfull{93} & \databarfull{88} & \databarfull{58} & \databarfull{53} & \databarfull{0} & \databarfull{17} & \databarfull{18} \\
    LeanPlum & 456 & 231 & \databarfull{18} & \databarfull{0} & \databarfull{60} & \databarfull{0} & \databarfull{98} & \databarfull{74} & \databarfull{13} & \databarfull{14} \\
    Gimbal & 396 & 160 & \databarfull{98} & \databarfull{82} & \databarfull{99} & \databarfull{81} & \databarfull{99} & \databarfull{58} & \databarfull{24} & \databarfull{3} \\
    Radius Networks & 369 & 94 & \databarfull{86} & \databarfull{83} & \databarfull{91} & \databarfull{0} & \databarfull{84} & \databarfull{0} & \databarfull{31} & \databarfull{21} \\
    Ad4Screen & 198 & 75 & \databarfull{0} & \databarfull{0} & \databarfull{0} & \databarfull{0} & \databarfull{0} & \databarfull{0} & \databarfull{8} & \databarfull{17} \\
    Kontakt & 195 & 82 & \databarfull{94} & \databarfull{91} & \databarfull{99} & \databarfull{0} & \databarfull{62} & \databarfull{0} & \databarfull{84} & \databarfull{3} \\
    Swrve & 153 & 76 & \databarfull{14} & \databarfull{0} & \databarfull{42} & \databarfull{0} & \databarfull{86} & \databarfull{0} & \databarfull{5} & \databarfull{25} \\
    Exponea & 99 & 68 & \databarfull{24} & \databarfull{3} & \databarfull{78} & \databarfull{3} & \databarfull{79} & \databarfull{3} & \databarfull{6} & \databarfull{50} \\
    Radar & 93 & 61 & \databarfull{38} & \databarfull{32} & \databarfull{99} & \databarfull{0} & \databarfull{80} & \databarfull{0} & \databarfull{25} & \databarfull{7} \\
    Bazaarvoice & 88 & 67 & \databarfull{38} & \databarfull{0} & \databarfull{95} & \databarfull{0} & \databarfull{88} & \databarfull{0} & \databarfull{19} & \databarfull{31} \\
    Rover SDK & 50 & 26 & \databarfull{98} & \databarfull{70} & \databarfull{94} & \databarfull{0} & \databarfull{84} & \databarfull{0} & \databarfull{12} & \databarfull{67} \\
    Zendrive & 34 & 16 & \databarfull{85} & \databarfull{59} & \databarfull{100} & \databarfull{0} & \databarfull{100} & \databarfull{79} & \databarfull{75} & \databarfull{8.30} \\
    Sensoro & 4 & 2 & \databarfull{100} & \databarfull{75} & \databarfull{100} & \databarfull{75} & \databarfull{75} & \databarfull{0} & \databarfull{100} & \databarfull{50} \\
    \bottomrule
    \end{tabular}
    }
    \parbox[t]{\textwidth}{\footnotesize \textit{Abbreviations Used:}
    \# Android 12+ - Number of apps targeting Android 12 or above; \\ 
    \% Req - Percentage of BLE\_SCAN permission requests; Man. - Manifest; \\ 
    API - API calls; \% Tag - Percentage of apps using the neverForLocation tag.
    }
    \label{tbl:beacon_perm_scan_stats}
\end{table}

We compare declared permissions in each app's
manifest with statically identified APIs to flag cases of over-permissioning.
Our analysis shows that 82\% of the apps request either fine or
coarse-grained location permissions, 
79\% of beacon-enabled apps request \wifi permissions and
62\% request Bluetooth-related permissions. This suggests the existence of close ties
between beacon SDKs and geolocation services. Interestingly, 28\% of
these apps request audio permissions, which in some cases could be used for
ultrasonic beaconing~\cite{mavroudis2017privacy, Arp2017privacy,
pfister2024listening}. 
However, the analysis of audio-tracking technologies is outside the scope of this paper.
Finally, over 40\% of apps request phone state permissions to access the IMEI on
Android 9 and 35\% the AAID on Android 12.

We inspect the code of apps requesting these permissions to
determine whether the caller class belongs to the app itself (first-party code) or the beacon
SDK (third-party code) using the backward slicing technique described in \S\ref{sec:methodology}.
This process allows us to infer the usage and purpose of API
invocations by reasoning about the business models of the SDKs requesting them
according to publicly available information.
In line with our static permission analysis, we find that API usage for location tracking
closely matches declared permissions. For \wifi, we observe an 11\% decrease in
API usage compared to declared permissions, suggesting possible
over-permissioning or the presence of obfuscated code. For Bluetooth, 18\% of apps invoke
Bluetooth APIs without declaring the corresponding permissions.
This under-permissioning could stem from cases like Facebook and Ogury Ad SDKs,
which provide multiple functionalities (i.e., advertising, location services,
BLE), while app developers may use them just for very specific purposes.

Meanwhile SDKs
like, AltBeacon and Estimote seem to have legitimate high rates of
Bluetooth-related permission requests---95\% and 97\%, respectively---for
detecting and interacting with beacons. Interestingly, 91\% of apps integrating
AltBeacon and also requesting location permission never invoke the corresponding
AltBeacon location APIs in Java code, indicating that the permission request
could be a requirement for complying with Android versions 11 and lower. On the
contrary, we observe how the majority of apps integrating Estimote, LeanPlum,
Gimbal or Radius Networks make use of the corresponding location APIs, as
described in \S\ref{sec:ecosystem}.

The dual purpose of
many Android permissions has been criticized for causing confusion among
developers and users. To illustrate this, 
we analyze the manifest files to extract the presence of
the \path{BLUETOOTH_SCAN} permission and the usage of the
\textit{neverForLocation} flag, introduced for apps targeting Android 12 or higher, as the official Android documentation
recommends~\cite{bleneverforlocation2024}. As summarized in Table
\ref{tbl:beacon_perm_scan_stats}, only 18\% of apps that include
AltBeacon and request Bluetooth scan permissions set also the
\textit{neverForLocation} flag, indicating that the remaining apps likely use
this permission both for BLE scanning and location purposes. Furthermore, only 3\%
of the apps integrating Kontakt SDK set this flag. Among the SDKs offering
location data aggregation---\eg Kochava, Adobe Experience Platform, and
Salesforce Marketing Cloud---the use of this flag ranges from 25\% to 35\%. We
observe that, while apps can use the \textit{neverForLocation} flag with the
\path{NEARBY_WIFI_SCAN} permission, only 11 apps request this permission, and
only 6 set the flag.

Overall, our results show that only a minority of apps using scan APIs
explicitly declare the intention of not to use these methods for location tracking
purposes through mechanisms like the \textit{neverForLocation} flag.
Meanwhile, a significant proportion of apps integrating SDKs like Gimbal,
LeanPlum, or Kochava appear over-permissioned, requesting more access than may
be necessary. Unfortunately, these practices can create potential opportunities
for third-party data harvesting, as we show next.

\subsection{Beacon Data Collection}
\label{subsec:access_beacon_collection}

\newcommand{\onlyandroidnine}{$\Diamond$}
\newcommand{\onlyandroidtwelve}{$\blacktriangledown$}
\newcommand{\bothandroid}{$\bigstar$}

\newcolumntype{x}[1]{%
>{\centering\hspace{0pt}}p{#1}}%

\begin{table}[tpb!]
    \centering
    \normalsize
	\caption{SDKs collect location data and exfiltrate identifiers classified
    under global persistent IDs, app persistent IDs, global resettable IDs, app
    resettable IDs, \wifi/Bluetooth scans, and GPS data (\S\ref{sec:threat_model}).
    Symbols indicate data collection on Android 9 (\onlyandroidnine), Android 12
    (\onlyandroidtwelve), or both (\bothandroid). Highlighted rows represent SDKs
    from our dataset; others are co-embedded SDKs detected through dynamic analysis.}
    \resizebox{.47\textwidth}{!}{%
\begin{tabular}{|l|r|p{1.4mm}p{1.4mm}p{1.4mm}p{1.4mm}p{1.4mm}p{1.4mm}|x{6mm}|p{2mm}p{2mm}|x{6mm}|p{1.4mm}p{1.4mm}p{1.4mm}p{1.4mm}p{1.4mm}p{1.4mm}|p{2mm}p{2mm}|}     
\hline
\multirow{2}{*}{\textbf{\makecell{\hspace{7mm}SDK \\\hspace{7mm}(\# of Apps)}}} & {} & \multicolumn{6}{c|}{\textbf{Glob. Pers.}} & \textbf{App. Pers.} & \multicolumn{2}{c|}{\textbf{\makecell{Glob.\\Rst.}}} & \textbf{App Rst.} & \multicolumn{6}{c|}{\textbf{\wifi/BLE scan}} & \multicolumn{2}{c|}{\textbf{GPS}} \\ \cline{3-20}
 & \textbf{\rotrot{\# App. Installs}} & \rotrot{Boot ID} & \rotrot{GSF ID} & \rotrot{IMEI} & \rotrot{HW ID} & \rotrot{\wifi MAC} & \rotrot{Email} & \rotrot{Android ID} & \rotrot{AAID} & \rotrot{BLE Name} & \rotrot{FID} & \rotrot{BLE ibeacon MAC} & \rotrot{ibeacon UUID} & \rotrot{Router MAC} & \rotrot{Router scan MAC} & \rotrot{Router Scan SSID} & \rotrot{Router SSID} & \rotrot{Coarse gelog.} & \rotrot{Fine geoloc.} \\ \hline
\rowcolor{orange!50} \textbf{Kochava} (220) & 2B & &  &  &  &  &  & \onlyandroidnine & \bothandroid & \bothandroid &  &  &  & \bothandroid &  &  & \bothandroid &  &  \\
\textbf{Yandex} (220) & 572M & &  & \onlyandroidnine &  & \bothandroid &  & \onlyandroidnine & \bothandroid &  &  &  &  & \bothandroid & \bothandroid & \bothandroid & \bothandroid &  &  \\
\textbf{Amplitude} (190) & 988M & &  & \onlyandroidnine &  &  & \onlyandroidnine & \onlyandroidnine & \bothandroid & \bothandroid &  &  &  &  &  &  &  & \bothandroid & \bothandroid \\
\textbf{Datadog} (140) & 146M & &  & \onlyandroidnine &  &  &  & \onlyandroidnine & \bothandroid &  & \bothandroid &  &  & \bothandroid &  &  & \bothandroid & \bothandroid & \bothandroid \\
\textbf{Sentry} (81) & 53M & &  &  &  & \onlyandroidnine &  & \onlyandroidnine & \bothandroid & \onlyandroidnine & \onlyandroidtwelve &  &  & \bothandroid &  &  & \bothandroid & \bothandroid & \bothandroid \\
\textbf{Omniture} (49) & 1B & &  &  &  & \onlyandroidnine &  & \onlyandroidnine & \bothandroid &  &  &  &  &  &  &  & \bothandroid & \bothandroid & \bothandroid \\
\textbf{Forter} (34) & 297M & &  & \onlyandroidnine & \onlyandroidnine &  &  & \onlyandroidnine &  & \onlyandroidnine &  &  &  &  &  & \bothandroid & \bothandroid &  &  \\
\rowcolor{orange!50} \textbf{Radar} (33) & 280M & &  &  &  &  &  & \onlyandroidnine & \onlyandroidnine &  &  &  & \bothandroid &  &  &  &  & \bothandroid & \bothandroid \\
\rowcolor{orange!50} \textbf{Huq Sourcekit} (24) & 25M & &  &  &  &  &  & \onlyandroidnine & \bothandroid & \bothandroid & \bothandroid &  &  & \onlyandroidnine &  &  & \onlyandroidnine & \bothandroid & \bothandroid \\
\textbf{cellrebel} (18) & 134M & &  &  &  &  &  &  &  &  &  &  &  & \bothandroid &  &  & \bothandroid & \bothandroid & \bothandroid \\
\textbf{Vizbee} (16) & 164M & &  &  &  &  &  & \onlyandroidnine & \bothandroid &  &  &  &  & \bothandroid &  &  & \bothandroid & \bothandroid & \bothandroid \\
\rowcolor{orange!50} \textbf{Cuebiq} (6) & 50M & &  &  &  &  &  &  & \bothandroid &  &  &  &  & \bothandroid & \bothandroid & \bothandroid & \bothandroid & \onlyandroidnine & \onlyandroidnine \\
\textbf{taobao} (6) & 1B & &  & \onlyandroidnine &  &  &  & \onlyandroidnine &  &  &  &  &  & \bothandroid &  &  &  &  &  \\
\textbf{My Tracker SDK} (6) & 132M & &  &  &  &  &  &  & \bothandroid &  &  &  &  & \bothandroid &  &  & \bothandroid &  &  \\
\textbf{AdsWizz} (6) & 9M & &  &  &  &  &  &  & \bothandroid &  &  &  &  &  &  &  & \onlyandroidnine & \bothandroid & \bothandroid \\
\textbf{phunware} (5) & 78K & &  &  &  &  &  &  & \bothandroid &  & \bothandroid &  &  & \onlyandroidnine &  &  & \bothandroid & \bothandroid & \bothandroid \\
\textbf{conviva} (5) & 134M & &  &  &  &  &  & \onlyandroidnine & \bothandroid &  &  &  &  & \bothandroid &  &  & \bothandroid & \bothandroid & \bothandroid \\
\textbf{PayPal} (5) & 100M & & \bothandroid & \onlyandroidnine & \onlyandroidnine &  &  & \onlyandroidnine &  &  &  &  &  & \bothandroid & \onlyandroidnine &  & \bothandroid & \bothandroid & \bothandroid \\
\rowcolor{orange!50} \textbf{Singlespot} (5) & 30M & &  &  &  &  &  &  & \bothandroid &  &  & \bothandroid &  &  &  &  &  &  &  \\
\textbf{Incognia} (5) & 38M & \bothandroid &  &  &  &  &  & \onlyandroidnine & \bothandroid &  &  &  &  & \bothandroid & \bothandroid & \bothandroid & \bothandroid & \bothandroid & \bothandroid \\
\rowcolor{orange!50} \textbf{Colocator} (4) & 395K & &  &  &  &  &  & \onlyandroidnine &  &  &  &  &  & \bothandroid &  &  &  &  &  \\
\rowcolor{orange!50} \textbf{Swrve} (3) & 9M & &  &  &  &  &  & \onlyandroidnine & \bothandroid &  &  &  &  &  &  &  &  &  &  \\
\textbf{JPush} (3) & 346K & &  &  &  &  &  &  &  &  &  &  &  &  &  & \bothandroid & \bothandroid &  &  \\
\rowcolor{orange!50} \textbf{Kontakt} (3) & 23K & &  &  &  &  &  &  &  &  &  & \bothandroid &  &  &  &  &  &  &  \\
\textbf{Proxy Cloud} (2) & 7M & &  &  &  &  &  &  &  &  &  &  &  & \bothandroid & \bothandroid & \bothandroid & \bothandroid & \bothandroid & \bothandroid \\
\textbf{pingID} (1) & 3M & &  &  &  & \onlyandroidnine &  & \onlyandroidnine &  &  &  &  &  &  &  & \onlyandroidtwelve & \bothandroid &  &  \\
\textbf{appICE} (1) & 3M & &  &  &  &  &  &  & \bothandroid &  &  &  &  & \bothandroid &  &  & \bothandroid & \onlyandroidnine & \onlyandroidnine \\
\textbf{Tangerine} (1) & 3M & &  &  &  &  &  &  & \bothandroid &  &  &  &  &  &  &  & \bothandroid &  &  \\
\textbf{Proximi.io} (1) & 162K & &  &  &  &  &  &  &  &  &  &  & \onlyandroidtwelve &  &  &  &  &  &  \\ \hline
\end{tabular}%
} 
\label{tab:locationidentifiers}
\end{table}

We study the runtime data access practices of beacon-enabled apps using the
dynamic analysis pipeline described in \S\ref{subsec:dynamic_analysis}. We
find that they collectively contact over 25K unique domains, with 86\% of apps
and 20\% of domains collecting at least one of the sensitive data types 
detailed in Table~\ref{tab:locationidentifiers}. While our 
dynamic analysis results are a lower-bound estimation of the potential practices
due to inherent coverage limitations of runtime analysis,
they provide actual evidence of SDK behaviors and potential privacy abuses.
Additionally, dynamic analysis extends SDK detection beyond static methods
by uncovering wireless scanning behaviors in advertising, analytics, 
and fraud-detection SDKs missed by static analysis due to code obfuscation
or reflection (\S\ref{subsec:limitations}).

We observe discrepancies between the app's runtime behavior and the declared safety
labels. 
We find that 2,292 apps collect
device identifiers without disclosing them. For location data, a different pattern
emerges: only 23\% of the 3,535 apps both declare the collection of location data and transmit it at runtime 
while 563 apps transmit location data
without disclosing it in their data safety labels.
For the rest of the analysis, we report how SDKs collect beacon and geolocation
data across apps and distinguish beacon SDKs from non-beacon SDKs that
also exhibit beacon scanning behaviors.

\vspace{1mm}
\parax{Router:}
SSIDs and BSSIDs represent the names and MAC addresses of connected and
nearby \wifi networks. They serve as precise location proxies~\cite{reardon201950}. 
We find that 1.95\% and 1.62\% of apps collect the
router SSID and BSSID of their connected \wifi access
points, respectively. 
At the SDK level, 29 of them collect either router SSIDs or
BSSIDs, with only 4 SDKs collecting both. 
Furthermore, 0.32\% of the apps and 8 different 
SDKs extend their reach to the SSID and BSSID of 
nearby \wifi access points.  
Leading beacon SDKs collecting AP data include Kochava, Colocator, Cuebiq---which
was removed from the Play Store due to invasive data collection
practices~\cite{Robertson2020}---and non-beacon SDKs like
Yandex, JPush~\cite{reardon2020jpush}, and Incognia that offer
advertising, fraud prevention, and push notifications
solutions~\cite{incognia_docs}. Jointly, these SDKs account for 4B total installs.

\vspace{1mm}
\parax{BLE:} we observe 218 apps disseminating the device's Bluetooth name---\ie
the user-friendly name often defined by users and potentially containing
PII like a user's name~\cite{detweiler2023bluetoothnaming}---to
cloud services. 
Additionally, 18 apps collect the AltBeacon UUID and 6 collect
the iBeacon UUID, both of them being unique and persistent IDs tied
to a specific physical location. At the SDK level,
7 beacon SDKs collect the device's Bluetooth address, 3 targets the iBeacon UUID, and 1
collects the iBeacon MAC address. Huq Sourcekit, Kontakt, and
Radar are the most prevalent SDK with such capabilities, 
with a cumulative install count of 280M users. These results suggest that
BLE scanning is comparatively less prevalent than \wifi scanning. This
discrepancy could stem from the more controlled scanning methods offered by
SDKs, or the use of geofencing---where scans are only enabled within
specific areas---that limit our dynamic testing approach, as detailed in \S\ref{subsec:limitations}.

\begin{figure}[!t]
    \centering
    \begin{lstlisting}[style=codesnip]
@com.geomobile.tiendeo outbound to api.radar.io:443@
POST /v1/logs HTTP/1.1
Content-Type: application/json
Host: api.radar.io
..
@"androidid":"XXXXXXXXXXXXXX"@
{
    "createdAt": 1720523270332,
    "level": "DEBUG",
    "message": "Ranged beacon | @beacon.type = IBEACON; beacon.uuid = 01022022-fa0f-0100-00ac-dd1c6502da1c; beacon.major = 53479; beacon.minor = 42571; beacon.rssi = -12"@
},
{
    "createdAt": 1720523270334,
    "level": "DEBUG",
    "message": "Handling beacon entry | @beacon.type = IBEACON; beacon.uuid = 01022022-fa0f-0100-00ac-dd1c6502da1c; beacon.major = 53479; beacon.minor = 42571; beacon.rssi = -12"@
},

"events":[], @"nearbyGeofences":[{"_id":"6318a0381c18820019e1e07e",@ "live":true, 
"type":"circle", "tag":"es.s.m","externalId": "103769","geometryCenter":
{"coordinates":[@longtitude: -X.XXXXX,latitude: XX.XXX@],"type":"Point"},    

\end{lstlisting}
\caption{iBeacon advertisements and geofence data exfiltrated to Radar.io,
including Android ID, beacon details (UUID, major, minor, RSSI), and geofence
metadata (coordinates and type).}
\label{fig:radar_leak}
\end{figure}

\vspace{1mm}
\parax{GPS Sensors}: 20\% and 18\% of apps actively collect coarse or
fine geolocation data, respectively. The majority 
of apps with beacon SDKs declare access to Android geolocation permissions, either fine
(78\%) or coarse location (77\%) and even when the app is put in the background (20\%).
The 13 beacon SDKs collecting location data include Radar, 
Salesforce Marketing Cloud, Rover SDK, LeanPlum, and Huq Sourcekit,
with a cumulative install count of 4B devices. 
Additionally, we observe 33 non-beacon SDKs within the same apps accessing the location
data, with OneSignal, Amplitude, Braze, 
and Flurry being the most prevalent ones. Among the SDKs that
collect the device's GPS location, 15 gather also the \wifi AP SSID and
BSSID. 
Similarly, as shown in Figure~\ref{fig:radar_leak}, Radar SDK collects
the device's GPS location along with BLE ibeacon data to potentially 
improve location accuracy and enable precise
location tracking~\cite{christl2024indoortracking}. Yet, this collection of 
location data solely for advertising or analytics purposes may potentially violate
Google's Play Store policies~\cite{googleplaypolicyads}.

\vspace{1mm}
\parax{Side-channels:}
we identify two non-beacon SDKs exploiting known vulnerabilities in older
Android versions to perform wireless scans by bypassing permission requirements
as of August 2024. The presence of such SDKs on the Play Store suggests that Google Play
Protect may fail to detect apps invoking vulnerable APIs.

\begin{itemize}[leftmargin=*]
\item \parax{Vizbee SDK} is a third-party library for mobile-to-TV
  deep-linking present in apps with over 164M installs.
  Vizbee exploits the side channel vulnerability 
  CVE-2020-0454~\cite{CVE-2020-0454} to collect AP's SSIDs
  in Android versions 9 and below. The SDK uses the
  \textit{onCapabilitiesChanged} callback function, registered via
  \textit{ConnectivityManager's} \textit{NetworkCallback}, which inadvertently
  provides SSID data. Vizbee caches and transmits this data to
  \path{metrics.clasptws.tv} under the key \texttt{WIFI\_SSID} (See Table~\ref{tab:vizbee} 
  in the Appendix). The transmission also
  includes \texttt{GEO\_LAT} and \texttt{GEO\_LONG} fields with values set to UNKNOWN, indicating
  that location permission is denied in this test. Vizbee stores the SSID in a variable named
  \textit{hackedSsid}, suggesting a deliberate attempt to bypass Android permission
  controls. 
  \item \parax{Forter SDK} (v2.4.11) present in apps with over 297M installs,
exploits the legacy API \textit{`WifiConfiguration.SSID`} in Android 9 devices to
collect SSIDs without requiring location permissions.
According to current market shares, these vulnerabilities remain active on approximately 9\% of
Android devices globally~\cite{androidvshare}. This SDK transmits sensitive data,
including SSIDs, security types (\eg WPA\_PSK), network associations, and device
IDs like the IMEI. In contrast, version 2.4.12 of the Forter SDK 
integrated in apps targeting Android 12
collects even more extensive network data, from 
DNS and DHCP details to \wifi scan results of nearby SSIDs and their
security configurations, device traffic statistics, device IDs, and
DNS settings.
\end{itemize}

\subsection{ID Bridging}
\label{subsec:id_bridging}

\begin{figure}[t!]
    \centering
    \includegraphics[width=\columnwidth]{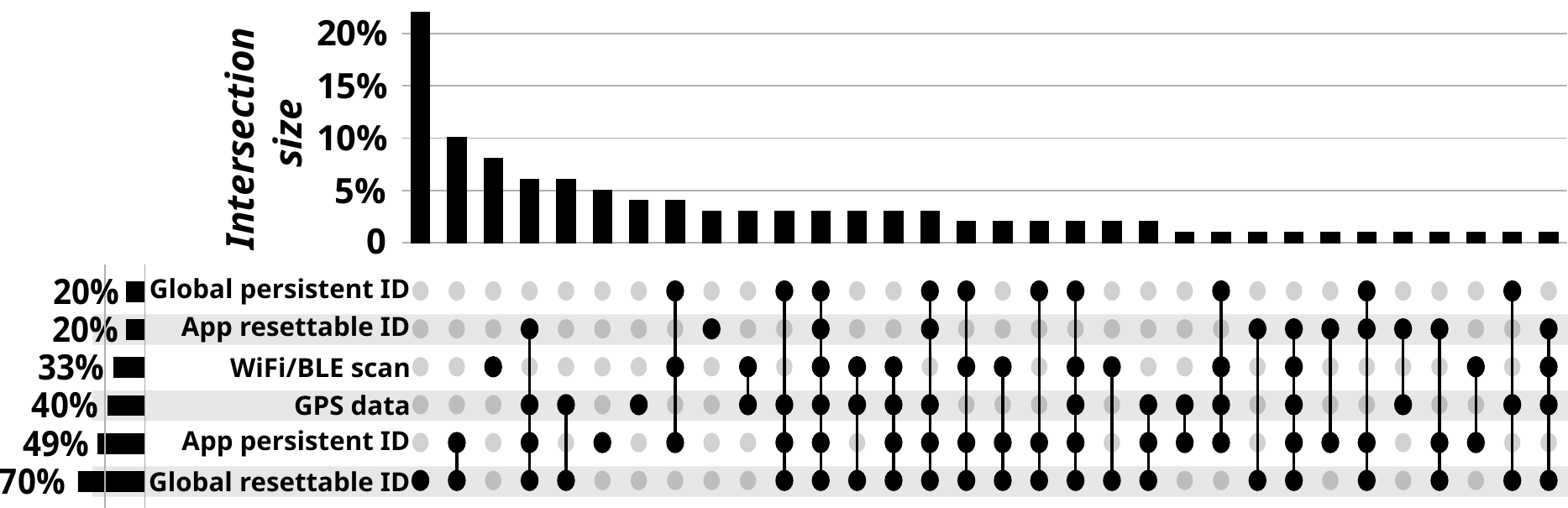}
    \caption{
    The UpSet plot illustrates how SDKs collect different combinations of ID
    categories. The top bars represent the percentage of SDKs collecting specific
    combinations (indicated by connected dots below), while the left bars show the
    total percentage of SDKs collecting each category, regardless of other data types
    collected.
  }
    \label{fig:upsetplot}
\end{figure} 

ID bridging is a privacy-intrusive practice that 
weakens the privacy protections of resettable identifiers (\eg AAID) and allows the 
creation of rich and persistent user profiles for advertising, identity 
profiling, and surveillance. 
We identify 61 SDKs potentially performing ID bridging, including 16 beacon SDKs %
and 45 non-beacon SDKs.
Figure~\ref{fig:upsetplot} summarizes the instances of ID bridging
captured by our dynamic analysis pipeline,
showing how beacon SDKs bridge
various types of device IDs, geolocation, and
BLE/\wifi scan results.

Overall, 70\% of the beacon SDKs also collect 
global resettable IDs like the AAID and BLE device name, 
and 41\% collect global persistent IDs like the GSF ID\@. 
Oftentimes, both persistent and resettable global IDs are 
bridged alongside \wifi or BLE scans (32\% of all SDKs) or geolocation 
coordinates directly collected from the GPS sensor (39\% of SDKs).
We note that the uniqueness of human mobility traces makes user de-anonymization
through geolocation data feasible, as demonstrated in prior
work~\cite{deMontjoye2013}. ID bridging strengthens this risk by linking
mobility data with other identifiers, allowing precise user re-identification
and persistent tracking.
Prominent beacon SDKs such as Kochava, Adobe Experience Platform, and Cuebiq, as
well as non-beacon SDKs present in the analyzed apps like Yandex, Adjust and
AdColony appear to perform these practices. We highlight some notable cases next.

\vspace{1mm}
\parax{WiFi/BLE Scans with Global IDs:} 
33\% of SDKs collect \wifi or Bluetooth scan data, 10\% combine
this data with global persistent IDs, and another 15\% with resettable IDs. 
These practices typically follow three main patterns:

\begin{table}[t!]
    \centering
    \large
    \caption{Packages with Yandex SDK, collecting multiple IDs.}
    \resizebox{\columnwidth}{!}{%
    \begin{tabular}{lp{6mm}p{2mm}p{2mm}p{2mm}p{2mm}p{2mm}p{2mm}p{2mm}p{2mm}p{2mm}}

        \toprule
        {} & {} & {} & {} & {} & {} & \multicolumn{4}{c}{\textbf{Router}}\\  
        \cline{7-10}

        \textbf{Package Name} & \rotrot{\textbf{Installs}} & \rotrot{\textbf{AAID}} & \rotrot{\textbf{Android ID}} & \rotrot{\textbf{WiFi MAC}} & \rotrot{\textbf{IMEI}} & \rotrot{\textbf{MAC}} & \rotrot{\textbf{SSID}} & \rotrot{\textbf{Scan MAC}} & \rotrot{\textbf{Scan SSID}} \\
        \midrule

                          ru.dvaberega &     968K &    x &            &         &      &         x &          x &               &                \\
                 com.numplates.nomera3 &     576K &    x &            &         &      &         x &          x &             x &              x \\
                 com.zenhotels.android &     261K &    x &          x &       x &    x &         x &          x &             x &              x \\
       com.wromanticgirlgame\_10378860 &      60K &    x &            &         &      &         x &          x &             x &              x \\
      com.weSPPTPBBKotaBogor\_13752947 &    39.4K &    x &            &         &      &         x &          x &             x &              x \\
                  com.ratehawk.android &    34.5K &    x &          x &       x &    x &         x &          x &             x &              x \\
       com.wMadaniQaidahUrdu\_15187981 &      515 &    x &            &         &      &         x &          x &             x &              x \\
              com.wBukuMenwa\_15656972 &      179 &    x &            &         &      &         x &          x &             x &              x \\
    com.AirportTransportation\_4990860 &       85 &    x &            &         &      &         x &          x &             x &              x \\

       \bottomrule
    \end{tabular}
    }
    \label{tab:yandex}
\end{table}

\begin{itemize}[leftmargin=*]
\item \parax{Connected Router Data:} 19\% of SDKs collect either resettable or
persistent user/device IDs along with connected \wifi network information (\eg
router MAC, router SSID), while 15\% of SDKs extend this by linking the Android
ID with connected router details. 
Namely, beacon SDKs such as Kochava, Adobe Experience Platform, and Colocator, along
with non-beacon SDKs like Conviva, PingID, AppICE, Taobao, Vizbee, and Datadog,
combine the AAID with the persistent GSF ID. For example, Figure~\ref{fig:kochava_leak} shows
Kochava collecting AAID with connected \wifi MAC. Similarly, non-beacon SDKs
like Yandex and Alipay incorporates the Android ID and \wifi MAC, enabling
cross-session tracking linked to precise network settings, regardless of AAID
resets. We observe these practices in 5\% of the apps that we analyze.
\item \parax{Nearby WiFi Scans:} Six SDKs upload nearby
    \wifi network data (\eg router scan SSID, router scan MAC), along with user
    IDs. For instance, Cuebiq and
    Incognia combine AAID and Android ID
    alongside both nearby and connected router scan data to track user movements
    across WiFi contexts, as shown in Figure~\ref{fig:incognia_leak}. 
    This data may be collected to enhance location services via crowd-sourcing.  
    Notably, Yandex combines AAID, Android ID, WiFi MAC,
    IMEI (in Android 9 and below), and even clipboard data with both connected and nearby router data. 
    Table~\ref{tab:yandex} lists popular apps with this behavior 
    integrating the Yandex SDK.

    \item \parax{BLE Beacon Data:} our analysis shows that three SDKs collect BLE
    beacon IDs such as iBeacon UUIDs or MAC addresses.
    This approach offers finer granularity than \wifi-based data and GPS
    location indoors. Proximi.io collects iBeacon UUIDs with device
    fingerprints, while the Kontakt SDK gathers iBeacon MAC addresses. Radar
    further integrates iBeacon UUIDs with GPS data, Android IDs, and AAIDs to
    enable detailed location-based tracking tied to physical BLE
    infrastructure even after AAID resets.

\end{itemize}

\vspace{1mm}
\parax{GPS Data with Global Resettable or Persistent IDs:} 39\% of SDKs collect
GPS data along with user IDs. This includes beacon SDKs such as LeanPlum, Huq
Sourcekit, and Cuebiq, and non-beacon SDKs like
Vizbee, Incognia, and Conviva.
Interestingly, Incognia goes a step further by harvesting the Boot ID, a 64-bit
unique hex string ID generated on a device's first boot and stored securely,
that should remain constant for the lifetime of the device unless the user
performs a factory reset.

\vspace{1mm}
\parax{AAID Bridging:} We find that 14\% of SDKs link the AAID
with global persistent IDs such as the IMEI (available only on Android 9 or below)
or GSF ID. In more than 180 apps, beacon-enabled analytics and marketing SDKs
such as Adobe Experience Platform and LeanPlum appear alongside non-beacon counterparts like
Amplitude, Sentry, and MixPanel, which appear to bridge the IMEI with the AAID\@. To protect
users' privacy, the Google Play Store user data policy explicitly prohibits linking
persistent IDs with resettable ones like AAID for advertising or analytics
purposes unless explicitly disclosed to users via privacy
policies or in-app consent dialogs~\cite{android_user_data_ids}.
However, our findings suggests that such linking still occurs, which could
potentially enable cross-app tracking, profiling, targeted advertising, and
third-party data aggregation for resale.
We also find non-beacon SDKs like MixPanel, Amplitude, and
New Relic bridging other IDs like the Android ID with the IMEI (in Android 9 and below) in 244
of the analyzed apps. MixPanel further enriches this profile by adding GSF ID.
The privacy implications of our findings are concerning. In fact, beacon SDKs
like Huq Sourcekit, Radar, Incognia, and Vizbee seem to 
opportunistically attempt to piggyback
on the permissions requested by the app developer to collect the full spectrum
of IDs along with GPS and BLE/\wifi data. 
For geolocation data, SDKs like Huq Sourcekit and Radar do not directly invoke location
APIs but instead appear to collect location data indirectly via host app permissions.
We demonstrate the risks of ID bridging with two case studies:

\begin{itemize}[leftmargin=*]
\item \parax{Amplitude:} This mobile analytics platform performs extensive data
 collection practices by bridging precise GPS coordinates with IDs like the
 AAID, the Android IDs, and the IMEI (in Android 9 and below). 
 Combining location data with persistent or resettable
 IDs for analytics purposes may violate Google's
 policy~\cite{googleplaypolicyads}.
 Additionally, Amplitude collects user email addresses, as seen in apps like
 \path{br.com.brainweb.ifood} (100M+ downloads), where emails are transmitted
 during the \textit{Sign In with Google} OAuth event triggered in our testing.
 Of particular concern is Amplitude's ability to
 track every user interaction within the app that embeds the SDK. Each
 user-triggered event is transmitted to Amplitude's servers, along with precise
 GPS coordinates, hence potentially enabling semi-continuous tracking of users'
 interactions and the exact locations where they occur.

 \item \parax{Adobe Experience Platform} collects various user identifiers through
 \textit{demdex.net} and \textit{*.omtrdc.net} domains, including resettable (AAIDs),
 globally persistent (WiFi MACs, hashed emails), geolocation (router SSIDs,
 GPS), and app- or device-scoped IDs (Firebase installation IDs, Android IDs).
 Additionally, Adobe sets its proprietary \textit{marketingCloudId}, a
 \textit{``persistent and universal identifier designed to track users across
 all Adobe Experience Cloud products and
 subsidiaries''}~\cite{adobe_target_identifying_visitors}. Such SDK-specific IDs
 add opacity, limit user control and enabling potential ID bridging.
 In over 16 apps, %
 we find the \textit{adobedtm} tracker to collect hashed emails alongside the
 \textit{marketingCloudId}. This data is stored in a dictionary
  labelled ``PII'' and logged under an event named \textit{PiiInformationReceived},
  suggesting an intentional collection of PII. We also
 observe cross-library interactions and potential data sharing. In
 \path{com.totalwine.app.store} (1M+ downloads), the \textit{marketingCloudId}
 is shared with Appsflyer alongside AAIDs, suggesting a direct interconnection
 with Adobe's identity graph solutions.

\end{itemize}

\subsection{Permission Usage Rationale}
\label{subsec:consent_rationale}

Android's official developer guidelines recommend that apps clearly communicate
the necessity of the permission requests and the potential impact if the user
denies it~\cite{permissionusage2024}. Android 6.0 introduced the
\textit{shouldShowRequestPermissionRationale()} method to inform
developers if their app should explain to the user the purpose of 
a requested dangerous permission.

Beacon SDK-enabled apps pose significant privacy risks to users while offering
them little to no control over data collection. To demonstrate this, we measure
their compliance with Android's permission request guidelines when requesting dangerous
permissions. We do this by statically searching for instances where
beacon SDKs invoke the \textit{shouldShowRequestPermissionRationale()} and
\textit{requestPermissions()} methods. We then parse the permission strings
associated with these function calls. Finally, we analyze the app's UI
components to detect any consent dialogs, such as alerts and banners, that might
be used to explain the permissions. By contextually linking all these elements,
we infer whether the app clearly provides a rationale for the requested
permissions.

We find that 24\% of apps in our dataset do not implement the
\textit{shouldShowRequestPermissionRationale()} API and include a rationale for permission
requests. Among those that do
implement the API, 92\% rely on third-party SDKs to provide such explanations
correctly, a transparency feature beyond the control of app developers. 
Additionally, the presence of these
justifications varies significantly across permission types: 71\% of apps
requesting location data fail to justify access,
only 13\% of apps justify \textit{FINE\_LOCATION} access, 7\% for
\textit{COARSE\_LOCATION}, and 8\% for \textit{BACKGROUND\_LOCATION}, in all
cases primarily handled by third-party SDKs. 
The lack of justification is even worse for Bluetooth permissions, with 97\% of
apps providing no explanation despite its known tracking risks. Our results
suggest that, most developers underestimate BLE data risks compared to location data.

On the other hand, only five beacon SDKs in our dataset use the 
\textit{shouldShowRequestPermissionRationale()} API to explain specific
dangerous permissions. %
Estimote shows rationales for coarse location in 75\% of apps embedding it.
Radar, Salesforce Marketing Cloud, and Singlespot show fine location rationales
in 42\%, 54\%, and 55\% of apps, respectively. Additionally, InMarket provides rationales for
accessing background location in 52\% of apps. 
In contrast, SDKs like Swrve and IndoorAtlas delegate this
responsibility entirely to app developers despite invoking
permission-protected APIs themselves. Swrve implements a callback to check if the rationale was shown to the
user~\cite{swrve_geo_sdk}, allowing it to piggyback on the app's permissions.

These findings highlight systemic transparency failures in how beacon-enabled
apps handle permission requests, leaving users with little insight or control
despite the dual usage of these permissions. Stricter auditing, policy
interventions and broader ecosystem reforms are necessary to address these gaps
as we discuss next.

\vspace{-1mm}
\section{Discussion}
\label{sec:discussion}

Our study reveals the pervasive tracking risks of beacon-enabled SDKs in
Android. By analyzing behaviors, cross-library interactions, and data collection
practices, we show how beacon SDKs track users across apps and services. Building
on Dehaye and Reardon's premise~\cite{dehaye2020proximity}, our findings
challenge the assumption that BLE-based distance authentication is secure
against global passive adversaries. The belief that large-scale Bluetooth
surveillance is economically unfeasible is flawed, as SDKs turn millions of
everyday apps into passive scanners. This shifts costs---power and
data collection---to users while enabling tracking of individuals who never
installed these apps. Their wireless devices, such as headphones or AirTags,
can be continuously observed by passive scanners. While our focus is on mobile
apps, similar tracking risks may exist---including connected
platforms (\eg IoT)---highlighting the need for better transparency, accountability, and
regulatory safeguards.

\parax{Beacon SDKs and Location Surveillance.} In \S\ref{sec:sdk_dissemination} we demonstrate how beacon SDKs continuously scan
for WiFi, Bluetooth, and GPS signals, %
supporting a vast data ecosystem of targeted advertising and data
brokers~\cite{keegan2021multibillion}. For instance, Singlespot, a French
marketing firm, claims to have collected data from 2 million users and sells it for up to
\$20,000 via platforms like Datarade~\cite{singlespot_profile}.  If persistent
identifiers are linked to specific geolocations, SDKs can then leverage WPS,
data brokers and public databases to infer the users' location and accurately
track their movements. Such knowledge is harmful to users' privacy because it
can disclose information about their personal beliefs or sexual orientation, or
it can even reveal classified information such as the location of secret
facilities~\cite{drakonakis2019please, zang2011anonymization, deMontjoye2013,
 wired2024_us-soldiers-spies-nuclear-germany}.
Beyond first-hand data collection, beacon SDKs may also share or repurpose the data
they obtained ~\cite{keegan2021multibillion}. As we analyzed in \S\ref{subsec:cross_library_analysis}, SDKs co-located on the same app can
exchange the collected information. This interconnectivity enables
pervasive data flows across organizations, where user-tagged beacon data spreads
across multiple third parties. %
Similarly, as shown in \S\ref{subsec:id_bridging}, proprietary identifiers like
Adobe's Marketing Cloud ID get linked with persistent and resettable user IDs
and then shared with other
SDKs. 
These practices circumvent user control and blur accountability.

\parax{Platform Policy Enforcement.}
Google Play has introduced policies designed to protect user privacy by
imposing strict rules on what data apps can collect and how they use it.
According to these policies, developers are responsible for ensuring that any
SDKs embedded in their apps do not sell or misuse sensitive user information
they collect~\cite{googleplay_user_data_policy}. 
For instance, one policy explicitly prohibits linking user data or resettable IDs (e.g., AAID) with
persistent device IDs or PII for advertising~\cite{googleplaypolicyads}. 
However, while analyzing apps in practice, we find that 14\% of SDKs
(\S~\ref{subsec:id_bridging}) bridge resettable identifiers with
persistent ones. Such identifier bridging undermines Android's privacy
safeguards, as resetting identifiers no longer limits continuous user tracking. 
A second recital mandates clear user disclosure for any location data collected
for ads, ensuring transparency and data minimization.  
Unfortunately, these policies shift the burden to developers. As they rush their
market release, regulatory complexity, inexperience, and incomplete SDK
documentation hinders compliance.  
To enforce compliance and detect policy violations, marketplaces could implement
stricter independent audits before app release, runtime monitoring, and publicly
disclose audit results for transparency.  

\parax{Regulation and Transparency.} Regulatory efforts in the EU and the US aim to
mitigate privacy risks and enhance transparency in mobile platforms.
GDPR~\cite{gdpr} mandates data minimization and consent, yet enforcement 
remains a challenge due to opaque SDK data-sharing practices. Meanwhile, CNIL and
AEPD have issued guidelines to curb passive wireless
tracking~\cite{aepd_wifi_tracking, cnil_mobile_apps_recommendation}. CNIL
advises app developers to map data flows, SDK providers to document compliance,
and marketplaces to enforce vetting. However, enforcement gaps persist as
widespread over-permissioning (\S\ref{sec:permission_analysis}), proprietary
identifier use by SDKs (\S\ref{subsec:id_bridging}), and uncontrolled SDK-driven
data sharing beyond user control enable large-scale tracking, making regulatory
oversight difficult. Our empirical findings reveal how beacon SDKs evade
existing protections through wireless scanning, cross-SDK interactions, and
non-compliant tracking. By highlighting these risks, our analysis calls for
strengthening audits, mandating SDK disclosures, and stricter enforcement 
to ensure compliance and curb unchecked data collection.

\parax{Defense Measures.} Our work highlights systemic transparency failures and
policy gaps in the data collection practices of beacon-enabled apps and Android's 
transparency features. Mitigating
these risks requires stronger privilege separation and sandboxing to limit
cross-library data sharing. While Google's Privacy Sandbox is in the right
direction, it only isolates advertising SDKs~\cite{GooglePrivacySandboxAndroid}.
Stricter runtime audits and app store vetting are needed to detect covert SDK
behavior, yet these rely solely on platform enforcement, leaving users with no
control over beacon data collection and sharing. This lack of control is
worsened by the absence of transparency mechanisms in the beacon ecosystem. As
shown in \S\ref{subsec:consent_rationale}, the
\textit{shouldShowRequestPermissionRationale()} API designed to improve user
awareness is rarely used, exposing gaps in permission governance. 
Existing privacy controls offer little protection against
abuse. The FCC suggests setting Bluetooth to hidden mode, but this only prevents
device discovery, not beacon scanning~\cite{fcc_protect_online}. Addressing
these risks requires a multi-layered approach, including stricter app store
policies, independent audits, and technical safeguards to curb unregulated
beacon tracking. While our work focuses on policy and platform-level privacy
controls, future research can explore usability improvements to enhance user
awareness and transparency.

\vspace{-1mm}
\section{Related work}
\label{sec:relwork}

\parax{\wifi and BLE Scanning.} 
Studies have shown how \wifi probe requests, which lack encryption and
authentication, are exploited to track users and establish social links using
SSIDs and BSSIDs~\cite{cunche2012know, freudiger2015talkative}. This data
contributes to large databases created through techniques like wardriving,
enabling passive location tracking by third
parties~\cite{sapiezynski2015opportunities}. Rye and Levin~\cite{rye2024surveilling}
highlighted the risks by revealing how Apple's \wifi Positioning System (WPS)
mapped geolocations for over 2 billion BSSIDs globally, enabling mass
surveillance. On the other hand, vulnerabilities in Bluetooth have been
exploited to track users~\cite{becker2019tracking, Das2016uncovering,
Kassema2021ble, Korolova2018cross, Kashi2020silentbeacons}, spoof, and perform
denial-of-service attacks~\cite{Kolias2017breaking} and exfiltrate private
data~\cite{priest2015covert}. 
Achara~\etal~\cite{achara2014short} highlighted how Android's \wifi permissions
were exploited by apps to infer user locations, prompting stricter controls by linking
\wifi data to location permissions. 

Several studies also identified privacy abuses across apps and SDKs. 
Reyes~\etal~\cite{reyes2018won} exposed
children's apps harvesting \wifi data to infer locations, violating COPPA.
Reardon~\etal~\cite{reardon201950} uncovered covert channels in Android used to
infer user locations. Dehaye and Reardon~\cite{dehaye2020proximity} showed SDKs
like X-Mode harvesting Bluetooth scans for user tracking to demonstrate the
existance of global passive adversaries in the context of contact tracing apps.
Building on this, our work provides the first empirical, large-scale analysis of
the Android beacon ecosystem, revealing how SDKs utilize GPS, BLE beacons, and
\wifi signals to track user proximity with high precision and link this data to
the user or device IDs.

\parax{Location.} Location privacy has drawn significant research attention due to
the rise of location-based services (LBS). A large body of work has focused on
GPS data dissemination in location-based social networks
(LBSNs)~\cite{zhao2016know, dhondt2022run, dhondt2024swipe, zhao2018you,
zhao2019geo}, crowdsourced location tracking~\cite{weller2020lost,
yu2024security, shafqat2023track, heinrich2021can}, and 
how GPS data from mobile apps~\cite{zhao2019geo, dhondt2022run, dhondt2024swipe}
inadvertently reveals sensitive information such as military base locations
or can be exploited for stalking and harassment~\cite{yu2024security, zhao2016know}.
On the other hand, studies of Call Detail Records (CDRs) and Twitter geolocation data revealed the
uniqueness and predictability of human mobility patterns~\cite{deMontjoye2013,
drakonakis2019please, gonzalez2008understanding}. 
De Montjoye~\etal~\cite{deMontjoye2013} demonstrated that with just four location
points it is possible to uniquely identify 95\% of individuals
However, there is a gap in the literature on how location data is
collected, who operates these location data services within the mobile
ecosystem, and how they can be misused to track users and their
movements.

\parax{Mobile App Privacy.} Privacy risks in mobile apps have been extensively
studied using static~\cite{nan2018finding, zhao2019geo, lu2023detecting,
octeau2013effective, arzt2014flowdroid, wei2018amandroid} and
dynamic~\cite{dhondt2024swipe, tileria2022watch, razaghpanah2018apps,
kollnig2021iphones, reardon201950} analysis techniques to uncover malicious
behaviors, third-party code implications~\cite{mahmud2022analysis,
xiao2024measuring, razaghpanah2018apps, feal2021don}, transparency
issues~\cite{MingASE2024, kollnig2021iphones,
arkalakis2024abandon}, and regulatory compliance gaps~\cite{koch2023ok,
kollnig2021fait, nguyen2021share, ZhaoICSE2023, nguyen2022freely}. 
However, the limitations of relying solely on static or
dynamic analysis~\cite{continella2017obfuscation, Choudhary2015ASE, GambaSP2020, poeplau2014execute} have led to hybrid
approaches that more effectively expose side and covert
channels~\cite{reardon201950, pourali2022hidden, wang2021understanding} and
privacy risks from analytical SDKs~\cite{demetriou2016free, razaghpanah2018apps,
son2016mobile, stevens2012investigating}.
Despite several proposed solutions like privacy policies and
labels, discrepancies still persist between stated and actual data usage
~\cite{arkalakis2024abandon, xiao2024measuring,
inayoshi2024detection}.
Our work does not use formal privacy accounting frameworks that rely on
mathematical models to estimate privacy issues, such as data de-anonymization
or re-identification~\cite{yu2024security, deMontjoye2013, dhondt2024swipe,
dhondt2022run}. On the contrary, we align with prior empirical privacy
research that combines static and dynamic analysis to measure the prevalence
of beacon data-collection behaviors and estimate the number of affected
users~\cite{reyes2018won, feal2021don, lyons2023log, reardon201950,
razaghpanah2018apps, nguyen2022freely, kollnig2021iphones}.

\vspace{-1mm}
\section{Conclusions}

This paper presents the first large-scale empirical analysis of
wireless-scanning SDKs in the Android ecosystem. By combining static and dynamic
analysis with signal injection and runtime monitoring, we reveal how 52 SDKs
across 9,976 apps exploit Bluetooth and WiFi scanning to infer user location and
collect sensitive data. Our findings reveal that this ecosystem is tightly
connected with advertising and tracking purposes and operates with minimal
oversight. Most SDKs collect geolocation data for such secondary purposes
and violate platform policies by engaging in ID bridging---linking persistent and
resettable identifiers to construct detailed user profiles without user consent
or knowledge for persistent user tracking. Some SDKs even intentionally exploit
side channels to access sensitive data and IDs without requesting the pertinent
Android permissions. We provide concrete evidence of non-compliance with
platform policies and inadequate enforcement of existing rules. Our study
demonstrates that existing privacy protections are insufficient, highlighting
the need for stricter regulatory control, robust SDK sandboxing to limit
cross-library data sharing, proactive audits to curb policy violations, and
stronger transparency mechanisms to prevent large-scale tracking and ensure
user control.

\begin{acks}
We thank the reviewers and the shepherd for their valuable feedback and suggestions
for improving our paper.
We used OpenAI's ChatGPT to correct grammatical errors and to improve the
clarity and coherence of our paper. The AI model was used with the following prompt: \textit{"please improve the
clarity and coherence of the text while maintaining the original intent"}.
This research was partially supported by project PID2022-143304OB-I00 (PARASITE) funded by MCIN/AEI
/10.13039/501100011033/ and by the ERDF “A way of making Europe”, 
the Spanish National Cybersecurity Institute (INCIBE) under
Proyectos Estratégicos de Ciberseguridad -- CIBERSEGURIDAD EINA UNIZAR,
and by the Recovery, Transformation and Resilience Plan funds,
financed by the European Union (Next Generation), the 
Natural Sciences and Engineering Research Council of Canada (NSERC)
(funding reference number RGPIN/04237-2018), and the Spanish AEI grant
CYCAD (PID2022-140126OB-I00).
Dr. Srdjan Matic was partially funded by the Atracción de Talento grant (Ref.
2020-T2/TIC-20184), funded by Madrid regional government.
Prof. N. Vallina-Rodriguez was appointed as 2019 Ramon y Cajal fellow (RYC2020-030316-I) 
funded by MCIN/AEI/10.13039/501100011033 and ESF Investing in your future. 
The opinions, findings,
and conclusions, or recommendations expressed are those of the authors and do
not necessarily reflect the views of any of the funding bodies.

\end{acks}

\bibliographystyle{ACM-Reference-Format}
\bibliography{main}

\appendix

\section*{Appendix}

\begin{table}[htpb!]
	\centering
  \scriptsize
	\caption{All 52 SDKs in our dataset and their purposes.}
	\resizebox{\columnwidth}{!}{%
		\begin{tabular}{lrccccccr}
			\toprule
			\textbf{SDK Name} & \textbf{\# Apps} & \multicolumn{6}{c}{\textbf{Purpose Type}}  \\
			\cmidrule(lr){5-8}
			                         &          &            &         & \rotrot{\textbf{Analytics}} & \rotrot{\textbf{Location}} & \rotrot{\textbf{Advertising}} & \rotrot{\textbf{Profiling}} \\
			\midrule
			AltBeacon                            & 4,022 &&  & \checkmark                  & \checkmark            &     &     \\
			Adobe Experience Platform                      & 1,328 &&  & \checkmark                  &                       &     &     \\
			Kochava                   & 1,117 &&  & \checkmark                  & \checkmark            &   \checkmark  &     \\
			Salesforce Marketing Cloud & 1,080 &&  & \checkmark                  & \checkmark            &   \checkmark  &     \\
			Estimote                             & 510   &&  & \checkmark                  & \checkmark            &     &     \\
			LeanPlum                 & 456   &&  & \checkmark                  & \checkmark            &     &   \checkmark  \\
			Gimbal                     & 396   &&  & \checkmark                  & \checkmark            &  \checkmark   &     \\
			Radius Networks            & 369   &&  & \checkmark                  &                       &  \checkmark   &     \\
			mParticle                            & 367   &&  & \checkmark                  &                       &     &     \\
			Ad4Screen                  & 198   &&  &                             &                       &   \checkmark  &     \\
			Kontakt                              & 194   &&  & \checkmark                  &                       &     &     \\
			CueAudio                             & 190   &&  & \checkmark                  & \checkmark            &     &     \\
			Swrve                     & 153   &&  & \checkmark                  & \checkmark            &     &  \checkmark   \\
			Reveal Mobile                        & 109   &&  & \checkmark                  & \checkmark            &     &     \\
			Exponea                              & 99    &&  & \checkmark                  &                       &     &     \\
			Radar                                & 93    &&  & \checkmark                  & \checkmark            &     &     \\
			IndoorAtlas                          & 92    &&  & \checkmark                  & \checkmark            &     &     \\
			SignalFrame                          & 89    &&  &                             & \checkmark            &     &     \\
			Bazaarvoice                          & 88    &&  & \checkmark                  & \checkmark            &     &     \\
			Huq Sourcekit                        & 81    &&  & \checkmark                  & \checkmark            &     &     \\
			Yinzcam Sobek              & 80    &&  & \checkmark                  & \checkmark            &  \checkmark   &     \\
			BlueKai (acquired by Oracle)         & 73    &&  & \checkmark                  &                       &     &     \\
			Cuebiq                               & 73    &&  & \checkmark                  & \checkmark            &     &     \\
			Rover SDK               & 50    &&  & \checkmark                  & \checkmark            &  \checkmark   &     \\
			Coulus Coelib              & 47    &&  &                             &                       &  \checkmark   &     \\
			Colocator                            & 40    &&  & \checkmark                  & \checkmark            &     &     \\
			X-Mode                               & 38    &&  & \checkmark                  & \checkmark            &     &     \\
			Zendrive                  & 34    &&  & \checkmark                  & \checkmark            &     &  \checkmark   \\
			Dynamic Yield                        & 27    &&  & \checkmark                  &                       &     &     \\
			Pilgrim by Foursquare                & 25    &&  &                             & \checkmark            &     &     \\
			Sense360                             & 24    &&  & \checkmark                  & \checkmark            &     &     \\
			Locuslabs                            & 23    &&  & \checkmark                  & \checkmark            &     &     \\
			InMarket                             & 23    &&  & \checkmark                  & \checkmark            &     &     \\
			Singlespot                           & 18    &&  & \checkmark                  &                       &     &     \\
			Roximity                             & 17    &&  &                             & \checkmark            &     &     \\
			Zapr                       & 17    &&  & \checkmark                  & \checkmark            &  \checkmark   &     \\
			Swirl                                & 16    &&  & \checkmark                  & \checkmark            &     &     \\
			Bluecats                             & 14    &&  &                             & \checkmark            &     &     \\
			Areametrics                          & 8     &&  & \checkmark                  &                       &     &     \\
			OpenLocate                           & 8     &&  &                             & \checkmark            &     &     \\
			Point Inside                         & 7     &&  & \checkmark                  & \checkmark            &     &     \\
			PredicIO                   & 6     &&  & \checkmark                  & \checkmark            &     &  \checkmark   \\
			MOCA                      & 5     &&  & \checkmark                  & \checkmark            &     & \checkmark    \\
			BeaconsInSpace (Fysical)             & 5     &&  &                             & \checkmark            &     &     \\
			Unacast Pure                         & 5     &&  &                             & \checkmark            &     &     \\
			Woosmap SDK                          & 5     &&  & \checkmark                  & \checkmark            &     &     \\
			Sensoro                              & 4     &&  & \checkmark                  & \checkmark            &     &     \\
			Signal360                            & 4     &&  & \checkmark                  &                       &     &     \\
			Placer                    & 4     &&  & \checkmark                  & \checkmark            &     &  \checkmark   \\
			Proximi.io                           & 3     &&  & \checkmark                  & \checkmark            &     &     \\
			Tamoco                               & 3     &&  & \checkmark                  & \checkmark            &     &     \\
			pulseid                              & 1     &&  & \checkmark                  & \checkmark            &     &     \\
			
			\bottomrule
		\end{tabular}
	}
	\label{table:full_tracker_list}
\end{table}

\begin{figure}[htbp!]
  \includegraphics[width=1.18\textwidth, angle=90]{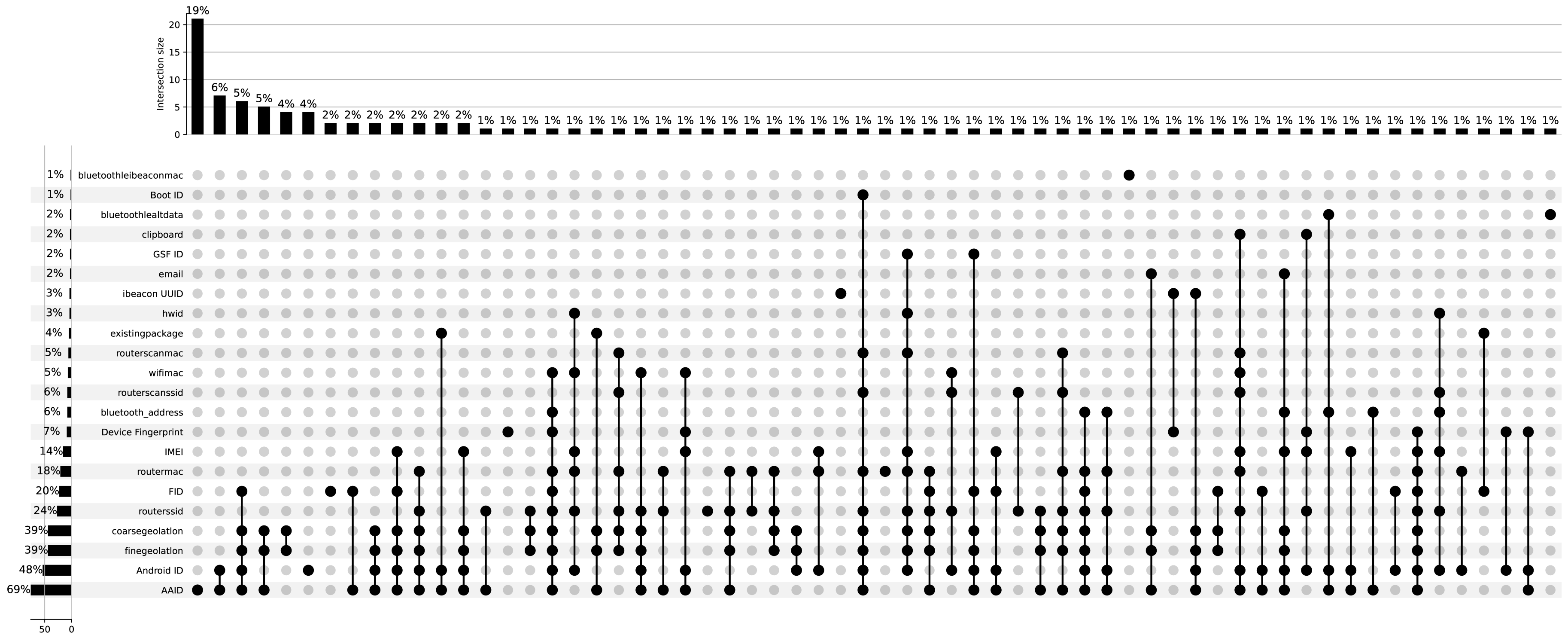}
  \caption{Upset plot of all the identifiers collected by the SDKs.}
    \label{fig:upset_plot_without_grouping_with_percentages}
\end{figure}

\begin{table*}[tpbh!]
    \small
    \caption{Description of PII types considered in this study. Abbreviations used: Pers - Persistent; Reset - Resettable.}
    \resizebox{\linewidth}{!}{%
    \begin{tabular}{lccl}
    \toprule 
    \textbf{Identifier}     &  \textbf{Pers}  & \textbf{Reset} & \textbf{Definitions}     \\ 
    \hline
    IMEI                    &            x          &                     & Unique hardware ID for the mobile device. Available only on Android 9 or below \\ 
    \wifi MAC               &            x          &                     & Unique hardware ID for the \wifi interface. \\ 
    HWID                    &            x          &                     & Unique hardware ID assigned by the manufacturer. \\ 
    GSF ID                  &            x          &                     & Unique ID tied to the Google account on the device. \\ 
    Boot ID                 &            x          &                     & Unique identifier representing the device's boot session; resets only after a factory reset. \\ 
    Email                   &            x          &                     & User's email address. \\ 
    Device Fingerprint      &                       &                     & Non-unique Collection of attributes (e.g., OS version, browser) identifying the device.. \\ 
    Android ID              &            x          &                     & App-specific ID that resets with a factory reset. \\ 
    AAID                    &                       &          x          & Advertising ID provided by Google for specialized for advertising. Can be reset through device settings. \\ 
    Firebase ID             &                       &          x          & App-specific ID tied to a specific app instance. It resets when the app is uninstalled or its data is cleared.  \\ 
    Bluetooth Device Name   &                       &          x          & User-defined, resettable, and non-unique name for Bluetooth discovery. \\ 
    Bluetooth iBeacon MAC   &            x          &                     & Unique hardware address for BLE proximity detection. \\ 
    iBeacon UUID            &            x          &                     & Unique ID for iBeacon devices in proximity-based services. \\ 
    Router MAC              &            x          &                     & Unique mac address for the connected \wifi router. \\ 
    Router Scan MAC         &            x          &                     & Persistent hardware address for nearby routers detected in \wifi scans, \\ 
    Router Scan SSID        &            x          &                     & Names of nearby routers detected in \wifi scans, \\ 
    Router SSID             &            x          &                     & Name of the currently connected router. \\ 
    Coarse Geolocation      &            x          &                     & Co-ordinates of approximate device location derived from network data. \\ 
    Fine Geolocation        &            x          &                     & Co-ordinates of precise GPS-based location of the device. \\ 
    \bottomrule
    \end{tabular}%
    } 
    \label{tab:pii_table}
    \end{table*}

\begin{table*}[h]
    \centering
    \small
    \caption{Example of Beacon SDK code and network signatures.}
    \begin{tabular}{lll}
        \toprule
        \textbf{SDK Name} & \textbf{Package Names} & \textbf{Domains / Endpoints} \\
        \hline
        AltBeacon & \makecell[l]{org.altbeacon.beacon., com.altbeacon.beacon., org.altbeacon.bluetooth.} & \makecell[l]{data.altbeacon.org} \\
        Radius Networks & \makecell[l]{{com.radiusnetworks.}} & \makecell[l]{{proximitykit.radiusnetworks.com}} \\
        Estimote & \makecell[l]{{com.estimote.}} & \makecell[l]{{*.estimote.com}} \\
        Gimbal & \makecell[l]{{com.gimbal.android.}} & \makecell[l]{{analytics*.gimbal.com}, {api.gimbal.com}, {sdk-info.gimbal.com}} \\
        Kontakt & \makecell[l]{{com.kontakt.sdk.android.}} & \makecell[l]{{kontakt.io}} \\
        Reveal Mobile & \makecell[l]{{com.stepleaderdigital.reveal.}} & \makecell[l]{{*.revealmobile.com} {rvl.wral.com}} \\
        SignalFrame & \makecell[l]{{com.wirelessregistry.observersdk.}} & \makecell[l]{{*.wirelessregistry.com}} \\
        IndoorAtlas & \makecell[l]{{com.indooratlas.android.sdk.}} & \makecell[l]{{ipsws.indooratlas.com}} \\
        Rover SDK & \makecell[l]{{io.rover.}} & \makecell[l]{{*.rover.io}} \\
        \bottomrule
    \end{tabular}
    \label{tab:sdk_signatures}
\end{table*}

\begin{table*}[h]
    \centering
    \small
    \caption{Beacon SDK search keywords and attribution signals.}
    \resizebox{\linewidth}{!}{
    \begin{tabular}{lll}
    \toprule
    \textbf{Technology} & \textbf{Search Keywords} & \textbf{Attribution Signals} \\ 
    \hline
    BLE & \makecell[l]{"beacon SDK", "proximity SDK", \\ "BLE SDK", "Bluetooth tracking SDK", \\ "BLE advertising SDK", "Bluetooth beacon SDK", \\ "RSSI scanning SDK"} 
        & \makecell[l]{Presence of BLE scanning APIs, \\ class names associated with BLE scanning, \\ declared BLE-related permissions in AndroidManifest.xml, \\ references to hardcoded UUIDs for beacon identification, \\ extraction of RSSI values for signal strength analysis} \\
    \hline
    WiFi & \makecell[l]{"WiFi beacon SDK", "proximity SDK", \\ "WiFi tracking SDK", "WiFi positioning SDK", \\ "WiFi scanning SDK"} 
        & \makecell[l]{Presence of WiFi scanning APIs, \\ class names related to WiFi scanning, \\ declared WiFi-related permissions in AndroidManifest.xml, \\ analysis of hardcoded SSIDs, references to known WiFi telemetry endpoints} \\
    \hline
    Location & \makecell[l]{"geofencing SDK", "proximity SDK", \\ "location tracking SDK", "indoor positioning SDK", \\ "RTLS SDK", "asset tracking SDK"} 
        & \makecell[l]{Presence of location-related APIs, \\ class names associated with location services, \\ declared location-related permissions in AndroidManifest.xml, \\ embedded geofencing configuration files, hardcoded location-based UUIDs} \\
    \bottomrule
    \end{tabular}
    }
    \label{tab:beacon_apis}
\end{table*}

\begin{table*}[h]
    \centering
    \small
    \caption{Android framework APIs (Java) that enable wireless beacon scanning.}
    \begin{tabular}{lll}
    \toprule
    \textbf{Technology} & \textbf{Permissions} & \textbf{APIs} \\ 
    \hline
    BLE & \makecell[l]{BLUETOOTH\_SCAN \\ BLUETOOTH\_CONNECT \\ ACCESS\_FINE\_LOCATION} 
        & \makecell[l]{android.bluetooth.BluetoothAdapter/startLeScan \\ 
                      android.bluetooth.BluetoothLeScanner/startScan \\ 
                      android.bluetooth.BluetoothDevice/connectGatt \\ 
                      android.bluetooth.BluetoothGatt/discoverServices \\ 
                      android.bluetooth.BluetoothLeAdvertiser/startAdvertising} \\
    \hline
    \wifi & \makecell[l]{CHANGE\_WIFI\_STATE \\ ACCESS\_WIFI\_STATE \\ ACCESS\_FINE\_LOCATION \\ ACCESS\_COARSE\_LOCATION}
        & \makecell[l]{android.net.wifi.WifiManager/getScanResults \\ 
                      android.net.wifi.WifiManager/startScan \\ 
                      android.net.wifi.WifiManager/getConnectionInfo \\ 
                      android.net.wifi.WifiManager/setWifiEnabled \\ 
                      android.net.wifi.WifiManager/reconnect} \\
    \hline
    Location & \makecell[l]{ACCESS\_FINE\_LOCATION \\ ACCESS\_COARSE\_LOCATION \\ FOREGROUND\_SERVICE}
        & \makecell[l]{android.location.LocationManager/getLastKnownLocation \\ 
                      android.location.LocationManager/requestLocationUpdates \\ 
                      android.telephony.TelephonyManager/getAllCellInfo \\ 
                      android.telephony.TelephonyManager/getCellLocation \\ 
                      android.location.LocationManager/addProximityAlert} \\
    \bottomrule
    \end{tabular}
    \label{tab:system_apis}
\end{table*}

\begin{table*}[h]
    \centering
    \LARGE
    \caption{Market share of the top-15 SDKs per top-10 app category.}
    \resizebox{\linewidth}{!}{%
    \begin{tabular}{lr|cccccccccc} %

        \toprule
     \textbf{SDK Name} & \textbf{\# Apps} &  \textbf{Lifestyle} &   \textbf{Shopping} &     \textbf{Sports} &   \textbf{Business} & \textbf{Travel \& Local} &   \textbf{Finance} & \textbf{News \& Magazines} & \textbf{Education} & \textbf{Entertainment} &  \textbf{Weather} \\
    \midrule    

    \textbf{AltBeacon                 } &            4024 & 18\% (721) &   6\% (240) &   3\% (137) &  12\% (498) &      8\% (338) &   3\% (133) &         1\% (59) &  8\% (302) &     3\% (114) &    1\% (32) \\
    \textbf{Adobe Experience Platform}  &            1328 &   3\% (45) &    7\% (89) &   9\% (120) &    4\% (47) &       5\% (65) &  13\% (179) &       16\% (212) &   1\% (13) &      7\% (97) &  18\% (245) \\
    \textbf{Kochava                   } &            1118 &   3\% (34) &    3\% (36) &    7\% (73) &    2\% (20) &       2\% (18) &    5\% (60) &         2\% (25) &   4\% (42) &      7\% (75) &     0\% (2) \\
    \textbf{Salesforce Marketing Cloud} &            1080 &   9\% (98) &  23\% (244) &    3\% (33) &  10\% (103) &       8\% (83) &  18\% (197) &         2\% (24) &   1\% (12) &      2\% (24) &     0\% (0) \\
    \textbf{Estimote                  } &             510 & 34\% (174) &    3\% (14) &     1\% (6) &   12\% (60) &      13\% (65) &     1\% (6) &          0\% (2) &  13\% (64) &      3\% (14) &     0\% (0) \\
    \textbf{LeanPlum                  } &             456 &   3\% (15) &    7\% (33) &     1\% (6) &    2\% (10) &      14\% (63) &    5\% (21) &          1\% (3) &  14\% (66) &       2\% (8) &     0\% (2) \\
    \textbf{Gimbal                    } &             396 &  10\% (40) &    6\% (23) &  42\% (167) &     1\% (5) &        1\% (5) &   20\% (80) &         3\% (10) &   5\% (21) &      6\% (24) &     0\% (0) \\
    \textbf{Radius Networks           } &             369 & 44\% (162) &    6\% (23) &     2\% (9) &    6\% (21) &       4\% (13) &     1\% (3) &          1\% (3) &   3\% (12) &       2\% (7) &     0\% (1) \\
    \textbf{mParticle                 } &             367 &   3\% (11) &    6\% (21) &    6\% (21) &    3\% (11) &      16\% (59) &    8\% (30) &        14\% (53) &    1\% (4) &      9\% (32) &     0\% (0) \\
    \textbf{Ad4Screen                 } &             198 &   5\% (10) &   19\% (38) &     3\% (5) &     2\% (3) &       7\% (14) &   12\% (23) &        29\% (58) &    1\% (2) &       2\% (3) &     0\% (0) \\
    \textbf{Kontakt                   } &             195 &   9\% (17) &   15\% (30) &    7\% (13) &   21\% (40) &       7\% (13) &     2\% (4) &          2\% (3) &    4\% (7) &       5\% (9) &     1\% (1) \\
    \textbf{CueAudio                  } &             190 &    0\% (0) &     0\% (0) &  98\% (187) &     0\% (0) &        0\% (0) &     0\% (0) &          0\% (0) &    0\% (0) &       1\% (2) &     0\% (0) \\
    \textbf{Swrve                     } &             153 &    2\% (3) &    8\% (13) &     3\% (4) &     0\% (0) &       7\% (11) &     5\% (8) &          1\% (2) &    2\% (3) &       5\% (7) &     1\% (1) \\
    \textbf{Reveal Mobile             } &             109 &    0\% (0) &     0\% (0) &     0\% (0) &     0\% (0) &        0\% (0) &     0\% (0) &        13\% (14) &    0\% (0) &       1\% (1) &   86\% (94) \\
    \textbf{Exponea                   } &              99 &    4\% (4) &   38\% (38) &     0\% (0) &     2\% (2) &      10\% (10) &   15\% (15) &          0\% (0) &    3\% (3) &       4\% (4) &     0\% (0) \\
    \midrule
    
    \textbf{Total  apps per category } &                   &  12\% (1385)&   9\% (999) &   8\% (935) &   8\% (903) &      7\% (847) &   7\% (795) &        5\% (600) &  5\% (562) &  4\% (473)  & 4\% (420) \\

    \bottomrule
    \end{tabular}
    }
    \label{table:market_share_table}
\end{table*}

\begin{table}[h]
    \centering
    \small
    \caption{Data collected by Vizbee.}
    \resizebox{\columnwidth}{!}{%
    \begin{tabular}{lcccccc}
        \toprule
                                              \textbf{Package Name} & \rotrot{\textbf{Installs}} & \rotrot{\textbf{AAID}} & \rotrot{\textbf{Router MAC}} & \rotrot{\textbf{Router SSID}} & \rotrot{\textbf{Android ID}} & \rotrot{\textbf{Coarse Geolocation}} \\
        \midrule
                                     com.gotv.nflgamecenter.us.lite &   120.1M &    x &         x &          x &          x &               x \\
                                     com.cnn.mobile.android.phone   &     50M  &    x &         x &          x &          x &                 \\
                                           com.handmark.sportcaster &    18.6M &    x &         x &          x &          x &                 \\
                                  com.turner.tnt.android.networkapp &     7.6M &    x &         x &          x &          x &                 \\
                                       com.nfl.fantasy.core.android &     7.4M &    x &         x &          x &          x &               x \\
                               com.neulion.android.tablet.nfl.wnfln &     3.9M &    x &         x &          x &          x &               x \\
                                                   com.turner.trutv &     1.7M &    x &         x &          x &          x &                 \\
                                                    com.fox.weather &   959.7K &    x &         x &          x &          x &               x \\
                                com.gannett.local.library.news.kare &   198.8K &    x &         x &          x &          x &               x \\
                                                    com.raycom.wtol &   194.1K &    x &         x &          x &          x &               x \\
                                com.gannett.local.library.news.wxia &   151.0K &    x &         x &          x &          x &               x \\
                                com.gannett.local.library.news.kxtv &    54.7K &    x &         x &          x &          x &               x \\
        \bottomrule
        \end{tabular}
    }
    \label{tab:vizbee}
\end{table}

\begin{table}[h]
    \centering
    \scriptsize
    \caption{SDKs that present rationale behind dangerous permission requested using shouldShowRequestPermissionRationale().}
    \label{tab:consent-rationale}
    \resizebox{\columnwidth}{!}{%
    \begin{tabular}{lcccc}
    \toprule
    \textbf{SDK} & \textbf{\% Ctx} & \textbf{C. Loc} & \textbf{F. Loc} & \textbf{B. Loc}  \\
    \midrule
    Estimote                   & \databarfull{75} & \ding{51} &                            &                            \\
    InMarket                   & \databarfull{52} &                            &                            & \ding{51} \\
    Radar                      & \databarfull{42} & \ding{51} & \ding{51} &                            \\
    Singlespot                 & \databarfull{55} &                            & \ding{51} &                            \\
    Salesforce Marketing Cloud & \databarfull{54} &                            & \ding{51} &                            \\
    \bottomrule
    \end{tabular}%
    } \parbox[t]{\columnwidth}{\footnotesize \textit{Abbreviations Used:} SDK -
    Software Development Kit, \% Ctx - \% of apps that Show Context, C. Loc -
    Access Coarse Location, F. Loc - Access Fine Location, B. Loc - Access
    Background Location.}
\end{table}

\begin{figure}[!t]
  \centering
  \begin{lstlisting}[style=codesnip, basicstyle=\small\ttfamily]
POST /track/json HTTP/1.1^M
User-Agent: Dalvik/2.1.0 (Linux; U; Android 12; AOSP on sargo Build/SP2A.220505.008)^M
Content-Type: application/json; charset=UTF-8^M
Host: control.kochava.com^M
Content-Length: 1729^M
^M

"action":"install","kochava_app_id":"kobiubiuclub-3b00","kochava_device_id":"KA31101723813375tdacae51f18644fed9ef9b8467b07d2be","sdk_protocol":"14","sdk_version":"AndroidTracker 3.11.0","nt_id":"bc770-1-acccf2f1-33da-4dbc-80bf-6a5981859f3f","data":{"screen_brightness":0.2902,"device_orientation":"portrait","volume":0.32,@"aaid": "X7dD24S2-CcD4-4Cfv-hhgg-bbbb7yk245f6"@,"device":"AOSP on sargo-Android","disp_h":2220,"disp_w":1080,"package":"com.biubiuclub.biubiuclubchat","installed_date":1723813319,"app_name":"BiubiuClub","app_version":"291","app_short_string":"2.9.1","os_version":"Android 

"network_conn_type":@"wifi","ssid":, "routerssid: ABC-WIFI", "bbsid": "routermac : XX:XX:XX:XX:XX:XX"@, "network_metered":false,"nvp":0,"carrier_name":"","usertime":1723813375,"uptime":0.945,"state_active":true,"app_limit_tracking":false,"platform":"android"},"sdk_id":"c946-s88450-","send_date":"2024-08-16T13:02:56.954Z"

\end{lstlisting}
  \caption{The Kochava SDK in \url{com.biubiuclub.biubiuclubchat} collects AAID and router MAC address and transmits them to control.kochava.com}
  \label{fig:kochava_leak}
\label{fig:kochava_leak}
\end{figure}

\begin{figure}[!t]
    \centering
    \large
    \begin{lstlisting}[style=codesnip, basicstyle=\small\ttfamily]
POST /events/v3 HTTP/1.1
x-dynatrace: MT_3_4_2021016446_1-0_36d6e79e-
"records": [
    {
      "app_session_id": "474c6944-823d-4da3-b154-001b272a6f59",
      "event_type": "bulk_localization_event",
      "localizations": [
        {
          "ad_tracking_enabled": "true",
          @"Ad_id": "ccccccc-cccc-cccc-cccc-ba1bccccc0a",@
          "app_package_name": "br.com.bancobmg.bancodigital.atletico",
          "app_state": "FOREGROUND",
          "fingerprint": {
            "elapsed_ts": "1589677460",
            "environment_scan": {
              "environment_state": "outdoor",
              "timestamp": "1723858094723"
            },

"gps_fix": {
                "altitude": "721.5999755859375",
                "bearing": "227.02573",
                "bearing_acc": "45",
                "gps_acc": "12.75",
                @"lat": "30.3368394",@
                @"lng": "-2.7705837",@
                "provider": "fused",
                "speed": "2.1049643",
                "speed_acc": "1.5",
                "vertical_acc": "1.5948421"
              },
              @"gps_ts": "1723858073227",@
              "mock_location": "false",
              @"precise_location": "true"@
            },
            "wifi_scan": {
              "ap_measures": [
                {
                  "ap_ts": "1723858075286",
                  "auth": "false",
                  @"bssid": "XX:XX:XX:XX:XX:XX",@
                  "channel_width": "20_mhz",
                  "con": "false",
                  "frequency": "2462",
                  "level": "-60",
                  @"ssid": "ABC-WIFII",@
                  "venue_name": "",
                  "wifi_rtt_responder": "false"
                },


\end{lstlisting}
    \caption{The Incognia SDK in \url{br.com.bancobmg.bancodigital.atletico} collects AAID, GPS coordinates, and WiFi scan data and transmits them to service2.us.incognia.com.}
    \label{fig:incognia_leak}
\end{figure}

\end{document}